\documentclass[AMA,STIX1COL]{WileyNJD-v2}
\usepackage{subcaption}
\usepackage{threeparttable}
\usepackage{listings}
\colorlet{punct}{red!60!black}
\definecolor{background}{HTML}{FFFFFF}
\definecolor{delim}{RGB}{20,105,176}
\colorlet{numb}{magenta!60!black}
\lstdefinelanguage{json}{
    numbers=left,
    numberstyle=\scriptsize,
    xleftmargin=2.2em,
    framexleftmargin=1.9em,
    stepnumber=1,
    numbersep=8pt,
    showstringspaces=false,
    breaklines=true,
    frame=single,
    backgroundcolor=\color{background},
    literate=
     *{0}{{{\color{numb}0}}}{1}
      {1}{{{\color{numb}1}}}{1}
      {2}{{{\color{numb}2}}}{1}
      {3}{{{\color{numb}3}}}{1}
      {4}{{{\color{numb}4}}}{1}
      {5}{{{\color{numb}5}}}{1}
      {6}{{{\color{numb}6}}}{1}
      {7}{{{\color{numb}7}}}{1}
      {8}{{{\color{numb}8}}}{1}
      {9}{{{\color{numb}9}}}{1}
      {:}{{{\color{punct}{:}}}}{1}
      {,}{{{\color{punct}{,}}}}{1}
      {\{}{{{\color{delim}{\{}}}}{1}
      {\}}{{{\color{delim}{\}}}}}{1}
      {[}{{{\color{delim}{[}}}}{1}
      {]}{{{\color{delim}{]}}}}{1},
}
\articletype{SPECIAL ISSUE PAPER}%

\received{26 April 2016}
\revised{6 June 2016}
\accepted{6 June 2016}

\usepackage{todonotes}

\begin{document}

\title{Function Delivery Network: Extending Serverless Computing for Heterogeneous Platforms }

\author[1]{Anshul Jindal}
\author[1]{Michael Gerndt}
\author[1]{Mohak Chadha}
\author[1]{Vladimir Podolskiy}
\author[2]{Pengfei Chen}

\authormark{Jindal \textsc{et al.}}

\address[1]{\orgdiv{Chair of Computer Architecture and Parallel  Systems}, \orgname{Technical University of Munich}, \orgaddress{\state{Garching (near Munich)}, \country{Germany\\}}}
\address[2]{\orgdiv{School of Data and
Computer Science}, \orgname{ Sun Yat-sen University}, \orgaddress{\state{Guangzhou}, \country{China}}}

\corres{Anshul Jindal, \\Chair of Computer Architecture and Parallel Systems, Technical University of Munich, Informatics 10, Boltzmannstr. 3 85748, \\Garching b. Munich Germany \\ \email{anshul.jindal@tum.de}}

% \presentaddress{This is sample for present address text this is sample for present address text}

\abstract[Summary]{Serverless computing has rapidly grown following the launch of Amazon's Lambda platform. Function-as-a-Service (FaaS) a key enabler of serverless computing allows an application to be decomposed into simple, standalone functions that are executed on a FaaS platform. The FaaS platform is responsible for deploying and facilitating resources to the functions. Several of today's cloud applications spread over heterogeneous connected computing resources and are highly dynamic in their structure and resource requirements. However, FaaS platforms are limited to homogeneous clusters and homogeneous functions and do not account for the data access behavior of functions before scheduling.

We introduce an extension of FaaS to heterogeneous clusters and to support heterogeneous functions through a network of distributed heterogeneous target platforms called Function Delivery Network (FDN). A target platform is a combination of a  cluster of homogeneous nodes and a FaaS platform on top of it. FDN provides Function-Delivery-as-a-Service (FDaaS), delivering the function to the right target platform.  We showcase the opportunities such as varied target platform's characteristics, possibility of collaborative execution between multiple target platforms, and localization of data that the FDN offers in fulfilling two objectives: Service Level Objective (SLO) requirements and energy efficiency when scheduling functions by evaluating over five distributed target platforms using the \textit{FDNInspector}, a tool developed by us for benchmarking distributed target platforms.  Scheduling functions on an edge target platform in our evaluation reduced the overall energy consumption by 17x without violating the SLO requirements in comparison to scheduling on a high-end target platform.}

\keywords{cloud computing, edge computing, high performance computing, serverless computing, function delivery network, function-as-a-service, heterogeneous platforms, heterogeneous faas}

\jnlcitation{\cname{%
 \author{Jindal A.}, 
 \author{M. Gerndt}, 
 \author{M. Chadha}, 
 \author{V. Podolskiy}, and 
 \author{P. Chen}} (\cyear{2020}), 
 \ctitle{Function Delivery Network: Extending Serverless Computing for Heterogeneous Platforms}, \cjournal{----.}, \cvol{---}.}

\maketitle

% \footnotetext{\textbf{Abbreviations:} ANA, anti-nuclear antibodies; APC, antigen-presenting cells; IRF, interferon regulatory factor}

\section{Introduction}\label{sec:introduction}

Presently, there exists a multitude of resources for processing and data storage ranging from small, inexpensive devices with limited computing resources to modestly priced servers with mid-range resources to
expensive high performance computers with extensive compute, storage, and network capabilities. These all combined form the \textit{computing continuum}~\cite{doi:10.1177/1094342019877383}. Many of today's applications are spread out over these heterogeneous connected computing continuum~\cite{10.1145/2391229.2391236}. (1) Web applications, for instance, combine mobile devices, edge computers for content delivery, and servers to enable interaction and collaboration. (2) IoT applications use micro-controllers, mini-computers, edge computers, and servers for delivering sensor measurements and controlling devices in the physical world. (3) Large scale experiments gather big data sets that need to be preprocessed and aggregated, forwarded to analytics functions, fed into compute-intensive simulations, and be visualized for the scientists. Many of these applications are highly dynamic with respect to their structure as well as the workload~\cite{10.1145/2391229.2391236}. Programming and deploying these applications is a highly challenging task. This is due to the heterogeneity of the underlying hardware, varying compute and data access requirements across time and application components, as well as the dynamic structure of the applications due to agile programming techniques combined with continuous delivery.  

% The reasons are the underlying heterogeneity of the hardware, the varying computes and data access requirements across application components, and across time, as well as the dynamic structure of the applications due to agile programming techniques combined with continuous delivery. 

Significant progress has been made in the context of cloud computing based on the idea of \textit{severless computing} since its launch by Amazon as AWS Lambda in November 2014~\cite{awslambdarelease}. Serverless computing is a cloud computing model that abstracts server management and infrastructure decisions away from the users~\cite{wg2018cncf}. In this model, the allocation of resources is managed by the cloud service provider rather than by the team of application developers and deployment managers, i.e., \textit{DevOps}, thereby increasing their productivity. Additionally, from the last couple of years there have been shift observed in the cloud native applications architecture from independently deployable microservices towards serverless architecture which is more decentralized and distributed~\cite{kratzke2018brief}.

% Serverless computing is a cloud computing model that abstracts server management and infrastructure decisions away from the users. In this model, the cloud service provider manages the allocation of resources instead of the team of application developers and deployment managers called \textit{DevOps} and thus increasing their productivity. 

% Function-as-a-Service (FaaS) is a key enabler of serverless computing. In FaaS, an application is decomposed into simple, standalone functions and these are uploaded to a FaaS platform for execution. These functions are stateless which means the state is not kept across function invocations. Functions can be invoked by a user’s HTTP request or another type of event created within the FaaS platform. FaaS platform is responsible for deploying and facilitating resources to the application functions.  

Function-as-a-Service (FaaS) is a key enabler of serverless computing~\cite{wg2018cncf}. In FaaS, an application is decomposed into simple, standalone functions that are uploaded to a FaaS platform for execution. These functions are stateless, i.e., the state is not kept across function invocations. Functions can be invoked by a user's HTTP request or by another type of event created within the FaaS platform. The FaaS platform is responsible for deploying and facilitating resources to the application functions.

% Currently, a significant number of open source and commercial FaaS platforms are available~\cite{faasplatforms}. All the large cloud providers have FaaS platforms available based on a \textit{container orchestration platform} such as Kubernetes. However, these FaaS platforms are limited to homogeneous clusters of nodes as well as to homogeneous functions. These assumptions facilitate the scheduling of functions invocations onto the available resources. Furthermore, FaaS platforms do not take the data access behavior of functions into account during scheduling. Since the functions are stateless, state changes and lookups require frequent access to databases which can result in data access latency.  
Currently, a significant number of open source and commercial FaaS platforms are available~\cite{faasplatforms}. All of the large cloud providers have FaaS platforms available based on a \textit{container orchestration platform} such as Kubernetes. However, these platforms are limited to homogeneous clusters of nodes as well as to homogeneous functions. These assumptions facilitate the scheduling of functions invocations onto the available resources. Furthermore, FaaS platforms do not account for the data access behavior of functions during scheduling~\cite{jonas2019cloud}. Since the functions are stateless, state changes and lookups require frequent access to databases that can lead latency in to data accesses.

In this article, we introduce an extension to the concept of FaaS as a programming interface for heterogeneous clusters and to support heterogeneous functions with varying computational and data requirements. This extension is a network of distributed heterogeneous target platforms called Function Delivery Network (FDN) analogous to Content Delivery Networks~\cite{gagliardi2011content}. A target platform is a combination of a cluster of homogeneous nodes and a FaaS platform on top of it. FDN provides Function Delivery as a Service (FDaaS), delivering the function to the right target platform based on the required computational and data demand. We target the integration of HPC clusters and distributed mini-computers (such as being used as edge devices) with the current platforms running on homogeneous clusters of servers in the cloud. In contrast to the elastic resource management in the cloud, HPC clusters are statically partitioned machines focusing on batch workloads. Space sharing is used to distribute the nodes to long-running applications that have exclusive access for their entire lifetime. The batch scheduling algorithm decides on the resource distribution to optimize the overall utilization of the system. Edge computers are currently used as a deployment device for a single application. In the IoT Greengrass system of Amazon~\cite{AWSIoTGr10:online}, it is already possible to integrate edge devices with cloud resources in an IoT platform and application Lambda functions running on it are deployed to the edge computers for implementing computing on the edge.
This approach is thus limited to single applications on the edge and a static distribution of computation. The integration of edge systems for general FaaS applications will require an extension of the FaaS platform across heterogeneous devices.

The automatic management of resources in the proposed serverless based FDN facilitates application development by shifting the burden to the cloud platform. However, already existing challenges like the fast startup of containers, communication, and latency of data accesses are further increased. The heterogeneity of the resources in the continuum is specifically challenging for resource management.
However, due to the heterogeneity of the FDN, it offers a wide range of opportunities for meeting different objectives like SLO requirements and energy efficiency in unconventional ways. Towards this, we present an external component of the FDN, \textit{FDNInspector}, a tool for benchmarking different target platforms and show based on our experiments the opportunities offered by FDN in meeting the two objectives: SLO requirements and energy efficiency. In summary, our main contributions are presented as follows.
\begin{enumerate}
    \item We propose an extension to the concept of FaaS as a programming interface for the computing continuum called Function Delivery Network (FDN). It should be noted that in this article we introduce the overall architecture of the FDN and describe its components in detail but the development of the FDN is still underway and therefore its implementation details are out of scope for this work. 
    \item We develop and present a tool called  \textit{FDNInspector} for evaluating distributed heterogeneous FaaS based target platforms. This tool is one of the external components of the FDN and is used for benchmarking the target platforms. It also contains the monitoring of the target platforms using Prometheus, which will be extended and reused as part of the FDN later.
    
    \item We highlight various opportunities offered by the FDN in meeting the two objectives: SLO requirements and energy efficiency when scheduling function invocations on five different target platforms by evaluating various function benchmarks using the developed \textit{FDNInspector}. 
    \item We present the performance evaluation results of target platforms for the introduced objectives and the opportunities provided by the FDN in meeting them.
\end{enumerate}

% The rest of this article is organized as follows. Section~\ref{sec:background} presents the basic background knowledge required for this article in brief. Section~\ref{sec:fdn} introduces the Function Delivery Network (FDN) and its components, in section~\ref{sec:methodology} we describe the overall experimental system design and the tool FDNInspector. Different goals and the performance evaluation results on those goals is presented in section~\ref{sec:results}.
The rest of this article is organized as follows. Section~\ref{sec:background} gives a brief overview of FaaS cloud model and the different FaaS platforms used in this work. In Section~\ref{sec:fdn}, the Function Delivery Network (FDN) and its components are introduced. Section~\ref{sec:methodology} describes our overall experimental system design and the tool \textit{FDNInspector}. Different goals and the performance evaluation results of the opportunities provided by the FDN in meeting those goals are presented in section~\ref{sec:results}. In Section~\ref{sec:discussion} a few additional opportunities provided by the FDN are discussed. In Section~\ref{sec:related_work}, we describe some of the previous works in this domain and in section~\ref{sec:threats_to_validity} we discuss the threats to validity.  Finally, Section~\ref{sec:conclusion} concludes the paper and presents an outlook.

\section{Background}\label{sec:background}
% This section introduces some of the concepts briefly which are used in this article. 
% \hl{In this section, we briefly describe the FaaS cloud model and }
In this section, we first present an overview of the FaaS cloud model. Following this, we describe the architecture and high level workflow of the three FaaS platforms used in this work.

\subsection{FaaS Cloud Model}\label{sec:faas_cloud_model}
Function-as-a-Service (FaaS) provides an attractive cloud model since it facilitates application development in which the user does not have to worry about the infrastructure management, but only about the code being deployed. The pricing is charged based on the number of requests to the functions and the duration, the time it takes for the function code to execute~\cite{AWSLambd84:online}. The latter varies according to the number of resources such as memory and CPU cores allocated to the function, and are automatically adapted to deliver the best performance. Instead of developing application logic in the form of services and managing the required resources, the application developer implements fine-grained functions connected in an event-driven application and deploys them into the FaaS platform~\cite{wg2018cncf}. The platform is responsible for providing resources for function invocations and performs automatic scaling depending on the workload.  The functions can be closely integrated with other services, e.g., cloud databases, authentication and authorization services, and messaging services. These services are called Backend-as-a-Service (BaaS). The Cloud Native Computing Foundation (CNCF) divides serverless into FaaS and BaaS~\cite{wg2018cncf}. BaaS are the third-party services that replace a subset of functionality in a function and allow the users to only focus on the application logic~\cite{lane2015overview}. In FaaS, function invocations are handled by using containers. Since functions are stateless, the state of the application is stored in databases. In comparison to microservice applications, FaaS has three advantages  (1) no continuously running services are required, (2) functions are only charged when they are executed, and (3) the function abstraction increases the developer's productivity.

One of the biggest differences between other forms of cloud models and the serverless model is scalability~\cite{scalabilityServerless}. In serverless computing, the application automatically scales up or down based on the resource usage (with scaling down to zero number of instances as well) and DevOps do not have to specify any scaling parameters. The infrastructure of the cloud service provider starts up ephemeral instances of each function on-demand. BaaS services are not set up to scale in this way unless the BaaS provider also offers serverless computing and the developers build this into their applications.

% Scalability is one of the biggest differentiators separating serverless model from other kinds of cloud models~\cite{scalabilityServerless}. In serverless computing, the application automatically scales up based on the usage. The cloud service provider infrastructure starts up ephemeral instances of each function on-demand. BaaS services are not set up to scale in this way unless the BaaS provider also offers serverless computing and the developer builds this into their application.

\subsection{FaaS Platforms}\label{sec:faas_cloud_platforms}
FaaS based functions can be invoked by a user's HTTP request or by another type of event created within the FaaS platform. The FaaS platform is responsible for providing resources for function invocations and performs automatic scaling. Currently a significant number of open source and commercial FaaS platforms are available~\cite{faasplatforms}. FaaS platforms implementations are based on starting containers for function invocations on top of a \textit{container orchestration platform} such as Kubernetes.  Applications are defined via a \textit{deployment specification} that describes the functions, APIs, permissions, configurations, and events that make up a serverless application. The specification can be given via a command-line or web interface, or by using some frameworks like Serverless~\cite{serverlessFramework} and Architect~\cite{architectFramework}. Updating of a deployment is also done through this deployment specification. All the updates in the specification are instantly propagated after which either the containers are restarted or only some configuration files are updated. 

% We introduce in the following subsections two open source FaaS platforms which are used as part of this work.

% FaaS platform is responsible for providing resources for function invocations and performs automatic scaling. FaaS platforms implementations are based on starting containers for function invocations on top of a container orchestration platform like Kubernetes. Currently a significant number of open source and commercial FaaS platforms are available~\cite{faasplatforms}. All the large cloud providers have FaaS platforms available based on a \textit{container orchestration platform} such as Kubernetes. Applications are defined via a \textit{deployment specification} that describes the functions, APIs, permissions, configurations, and events that make up a serverless application. The specification can be given via a CLI or web interface or using some frameworks like Serverless~\cite{serverlessFramework} and Architect~\cite{architectFramework} . Besides initial deployment, updating a deployment is also done through a deployment specification. All the updates in the specification are instantly propagated after which either a container is restarted or only some configuration files are updated. We introduce in the following subsections two open source FaaS platforms which are used as part of this work.

\begin{figure}[t]
\centering
\includegraphics[width=0.65\linewidth]{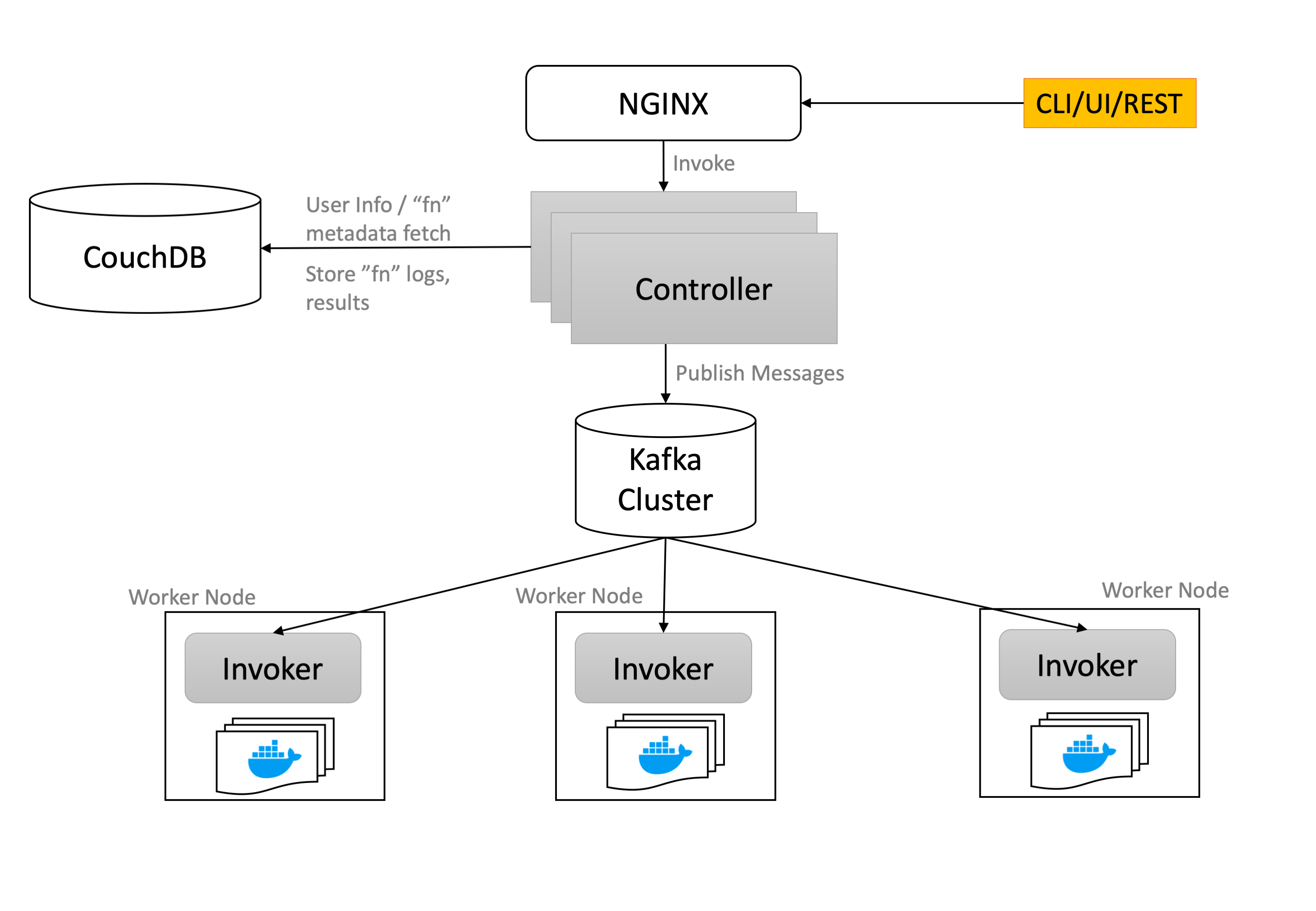}
\caption{\label{fig:openwhisk_archi}Openwhisk high level workflow~\cite{openwhiskDocs}}
\end{figure}

\subsubsection{OpenWhisk}
\label{sec:open_whisk}

% Apache OpenWhisk is a serverless open source cloud platform which was originally developed by a research group at IBM in 2015 and was released in december 2016 but after that it was donated to the Apache Software Foundation~\cite{pierre2016openwhisk}. It powers IBM's serverless offering, IBM Cloud Functions. It implements FaaS on top of the Kubernetes as the container orchestration platform. Functions in OpenWhisk are called actions and the execution of an action is called an
% invocation. Actions and rules can be created through their command line interface (wsk~\cite{wskCLI}), user interface or SDK. Created actions can then be invoked either manually through the same methods or by event triggers. Events can originate from multiple sources, including timers, databases, message queues, or websites like Slack or GitHub. 
Apache OpenWhisk is a serverless open source cloud platform that was originally developed by a research group at IBM in 2015 and was released in December 2016. It was later donated to the Apache Software Foundation~\cite{pierre2016openwhisk}. It powers IBM's serverless offering, IBM Cloud Functions and implements FaaS on top of Kubernetes as the container orchestration platform. Functions in OpenWhisk are called actions and the execution of an action is called an invocation. Actions and rules can be created through the command-line interface (CLI) (\texttt{wsk}~\cite{wskCLI}), user interface (UI), or SDK. Created actions can then be invoked either manually through the same methods or by event triggers. Events can originate from multiple sources including timers, databases, message queues, or websites like Slack or GitHub. 

% Actions and rules can be created through their user or command-line interface (\texttt{wsk}~\cite{wskCLI}), or by using the SDK.

OpenWhisk consists of multiple components under the hood as shown in the Figure~\ref{fig:openwhisk_archi} and all the components are packaged inside their individual docker containers when OpenWhisk is deployed~\cite{openwhiskDocs}. Each function invocation is translated into an HTTP request to the Nginx server~\cite{nginx}. The Nginx server is a single point of entry and its main purpose is to implement the support for the HTTPS secure web protocol. On receiving a request, the Nginx server forwards it to the controller. The controller is responsible for authenticating and authorizing the requests in coordination with CouchDB where all the user's data and their privilege levels are stored. The controller also has a load balancer which keeps track of the availability of the invokers, i.e., the workers that run the code and chooses one of them for the invocation. Controller and invokers communicate through Kafka~\cite{garg2013apache}, a publish-subscribe messaging system. The controller publishes the messages to Kafka addressed at a chosen invoker and once the message delivery is confirmed by the invoker, an HTTP request is sent back to the user with an \textit{ActivationId}, which can be used for retrieving the results of this function call. This processing is asynchronous, however synchronous processing is also available. It functions similarly to asynchronous processing, except in this case, the client will block until the action is completed and will retrieve the results immediately. Invokers set up a new docker container for each action, inject the code into them, execute the code, obtain the results, and then destroy it. These containers are run inside Kubernetes pods. There can be an invoker per kubernetes worker node as shown in the Figure~\ref{fig:openwhisk_archi} or an invoker can be responsible for managing multiple kubernetes worker nodes. Functions can also be chained together into sequences where chained functions use the output of the preceding function as input. OpenWhisk supports running functions in languages: 
Python, Node.js, Scala, Java, Go, Ruby, Swift, PHP, Ballerina, .NET and Rust~\cite{openwhisk2018apache}. Functions which are not using these languages can be created by providing a custom built docker runtime.

\subsubsection{OpenFaaS}
\label{sec:open_faas}

OpenFaaS is an another widely popular open source serverless cloud platform hosted by OpenFaaS Ltd~\cite{openfaas_archi}. Until March 2019, it was developed by a team of full-time developers from VMWare~\cite{openfaas_movement}. It also implements FaaS on top of Kubernetes as the container orchestration platform. Functions in OpenFaaS can be written in any language, and unlike  OpenWhisk, one does not have to create custom runtimes to make it work. A pre-built docker image of the function can be supplied to it. 

\begin{figure}[t]
\centering
\includegraphics[width=0.65\linewidth]{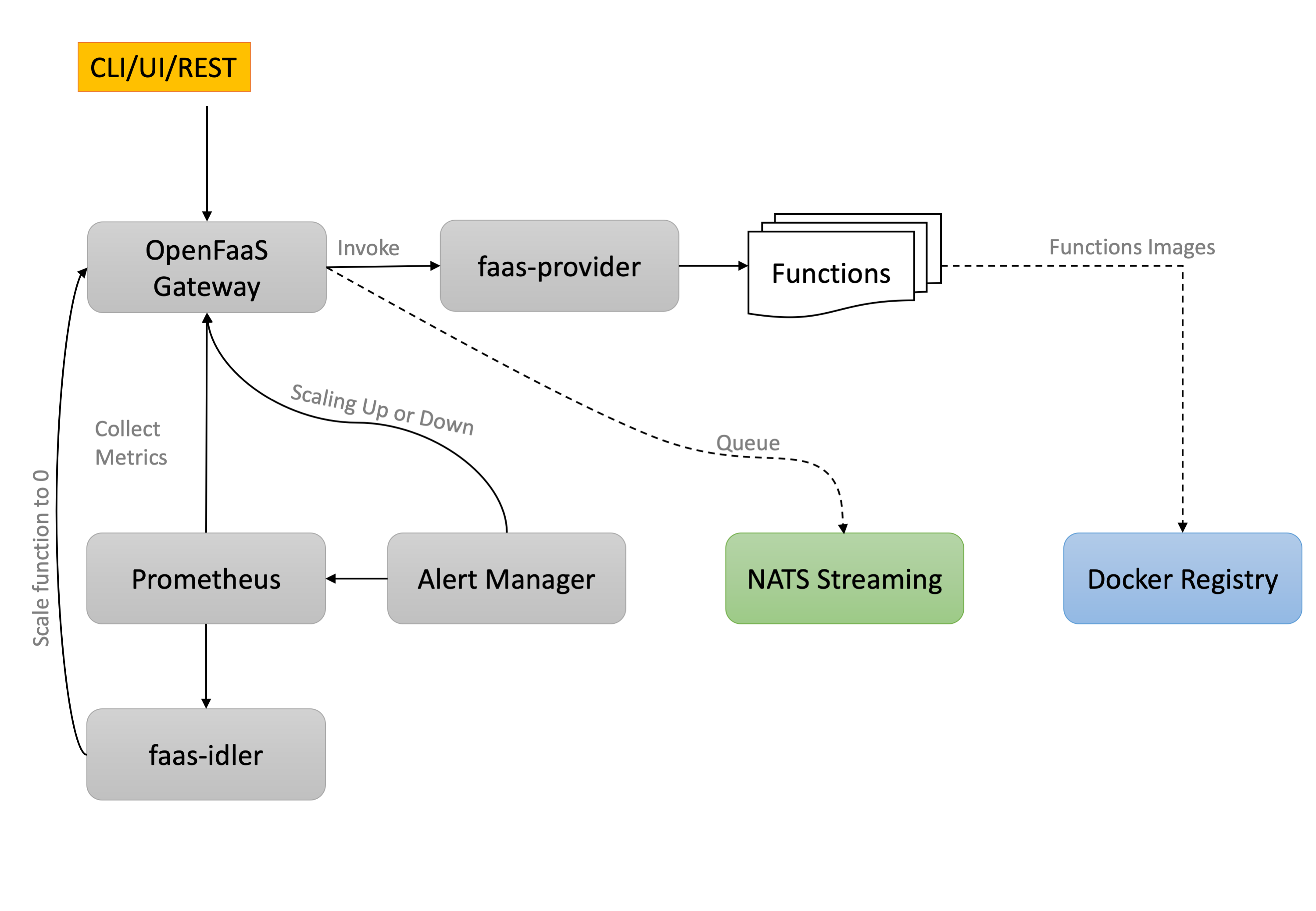}
\caption{\label{fig:openfaas_high_level_workflow}OpenFaas high level workflow~\cite{openfaas_archi}.}
\end{figure}

% Function deployment can happen similarly like OpenWhisk through any interface to the OpenFaaS Gateway (CLI/UI/REST), either manually or by setting up triggers. 
Similar to OpenWhisk, functions can be deployed through any interface to the OpenFaaS Gateway (CLI/UI/REST), either manually or by setting up triggers. OpenFaaS Gateway is the single point of entry for all the requests.  Figure~\ref{fig:openfaas_high_level_workflow} shows a high level workflow of the interaction between the different components of OpenFaaS. From the gateway, CRUD (create, read, update, delete) operations and invocations are forwarded to the \textit{faas-provider}, i.e., the controller which translates OpenFaaS functionality to a certain provider. \textit{faas-netes}~\cite{faas_netes} is an example of a faas-provider in OpenFaaS which enables Kubernetes for it. Because of this transparency to Kubernetes, one can interact with OpenFaaS resources directly through \textit{kubectl}, the command line interface for Kubernetes. When a function is created, its code is pulled from the docker registry and executed inside a container. It utilizes \textit{Prometheus} and its \textit{AlertManager} to continuously expose metrics. The AlertManager uses these metrics to determine auto-scaling decisions and inform them to the OpenFaaS gateway which then scales the function replicas up or down. The minimum (initial) and maximum replica count can be set at the time of deployment by adding a label to the function. When using Kubernetes, the built-in Horizontal Pod Autoscaler (HPA) can also be used instead of AlertManager~\cite{hpa_openfaas}. Scaling to zero to recover idle resources is available in OpenFaaS, but is not turned on by default. Scaling down to zero replicas is also called "idling" in OpenFaaS. The \textit{faas-idler}, an external component is responsible for making the scaling down to zero decision~\cite{faas_idler}. It monitors the built-in Prometheus metrics on a regular basis along with the \texttt{inactivity\_duration} variable to determine if a function should be scaled to zero or not. Only functions with a label of \texttt{com.openfaas.scale.zero=true} are scaled to zero, all others are ignored. When using \textit{faas-netes} as the provider, \textit{faas-idler} is automatically deployed by default.

OpenFaaS's watchdog is responsible for starting and monitoring functions in OpenFaaS~\cite{openfaas_watchdog}. It provides a generic interface between the outside environment and the function. The watchdog is a tiny Golang webserver which every function uses as their docker \texttt{ENTRYPOINT}. It acts as the init process for function container. Once the function is invoked, the watchdog passes in the HTTP request via \texttt{stdin} and reads a HTTP response via \texttt{stdout} and sends it back to user. 

OpenFaaS enables long-running tasks or function invocations to run in the background through the use of NATS Streaming~\cite{t2019study}. This decouples the HTTP transaction between the caller and the function. The HTTP request is serialized to NATS Streaming through the gateway as a "producer". The queue-worker acts as a subscriber and deserializes the HTTP request and uses it to invoke the function directly.  To fetch the results from an asynchronous call, the user can specify a callback url.

\subsubsection{Google Cloud Functions (GCF)}
\label{sec:gcf}
Google Cloud Functions is a serverless execution environment for building and connecting services in a cloud-based application offered by Google Compute Platform(GCP)~\cite{GoogleCloudFunctions:online}. With Google Cloud Functions, developers do not need to provision any infrastructure or worry about managing any servers, the whole environment including infrastructure, operating systems, and runtime environments are managed by Google. Currently, Cloud Functions supports JavaScript, Python 3, Go, and Java runtimes. Cloud Functions are simple, single-purpose functions that are attached to events emitted from the cloud infrastructure and services. The function is triggered when an event being watched is execcuted.  These events can be things like changes in a database, files added to a storage system, or a new virtual machine instance is created. A response to an event is created using a trigger which can then be attached to a function to capture and act on events. GCFs can either be deployed using the web interface or the gcloud\footnote{https://cloud.google.com/sdk/gcloud} command line tool.

Each Cloud Function runs in its own isolated secure execution context, scales automatically, and has a lifecycle independent from other functions~\cite{GoogleCloudFunctionsExec:online}. Cloud Functions handles incoming requests by assigning them to instances of function. Depending on the volume of requests, as well as the number of existing function instances, Cloud Functions may assign a request to an existing instance or create a new one. Each instance of a function handles only one concurrent request at a time. Thus the original request can use the full amount of resources (CPU and memory) that is requested. In cases where inbound request volume exceeds the number of existing instances, Cloud Functions start multiple new instances to handle requests. This automatic scaling behavior allows Cloud Functions to handle many requests in parallel, each using a different instance of the function.

%\subsection{Computing Continuum}\label{sec:computing_continuum}
%Currently, there exists a multitude of resources for processing and data storage such as HPC clusters, servers, workstations, mobile devices, mini computers, and microcontrollers. These all combined together are called as \textit{computing continuum}. 

%Many of today's applications are spread out over these heterogeneous connected computing resources. (1) Web applications, for example, combine mobile devices, edge computers for content delivery, and servers to enable interaction and collaboration. (2) IoT applications use microcontrollers, mini computers, edge computers, and servers for delivering sensor measurements and controlling devices in the physical world. (3) Large scale experiments gather big data sets that need to be preprocessed and aggregated, forwarded to analytics functions, fed into compute intensive simulations, and be visualized for the scientists. Many of these applications are highly dynamic with respect to the application structure as well as the load.  Programming and deploying those applications is a highly challenging task. The reasons are the underlying heterogeneity of the hardware, the varying compute and data access requirements across application components and across time, as well as the dynamic structure of the applications due to agile programming techniques combined with continuous delivery.
%Applications on the computing continuum use instead a federation of heterogeneous resources, called \textit{target platforms} in this proposal.

\section{Function Delivery Network (FDN)}\label{sec:fdn}

% Serverless computing in the form of FaaS is extremely attractive to DevOps as they are released from managing resources and from autoscaling application's components. 

Serverless computing in the form of FaaS is extremely attractive to DevOps as they are no longer responsible for managing infrastructure resources and autoscaling application components. FaaS provides automatic scaling for each function invocation as a result of a trigger. These invocations are then automatically distributed across the available resources. Current FaaS platforms are limited to clusters of homogeneous nodes.  However, many cloud applications in the computing continuum require heterogeneous resources for the execution. At a high level, heterogeneity in FaaS exists in two ways: 
\begin{itemize}
     \item One by using FaaS over heterogeneous clusters,  clusters  with different system architectures. For example, one cluster consisting of VMs in the Cloud, and another cluster consisting of resource-constrained edge devices. Such a method has an advantage of achieving higher application performance by placing the functions into the specific clusters depending on their computational requirements, and could even be used for reduction in the overall energy consumption~\cite{beloglazov2012energy}.

    \item Second by using heterogeneous FaaS platforms. Due to resource constraints in edge devices not all serverless platforms can run on them. In~\cite{8817155} four open source serverless frameworks, namely, Kubeless, Apache OpenWhisk, OpenFaaS, Knative are evaluated on resource-constrained edge devices. Also, Pfandzelter et al.~\cite{pfandzelter2020tinyfaas} highlight the problem of running cloud based FaaS platforms on the edge and introduce a new FaaS platform called tinyFaaS for edge environments. Therefore, one cannot run a homogeneous FaaS platform over heterogeneous clusters. 

\end{itemize}
In this article,  a target platform is a combination of a homogeneous cluster and a FaaS platform on top of it. For extending the serverless computing FaaS platform to heterogeneous clusters and to support heterogeneous functions with varying computational and data requirements we introduce a network of distributed target platforms called as \textit{Function Delivery Network (FDN)} analogous to Content Delivery Networks distributing web content and media to a network of distributed resources to provide service with the best quality of service (QoS). The FDN provides \textit{Function Delivery as a Service (FDaaS)}, delivering the function to the right target platform based on the computational and data requirement. 

When extending FaaS to target platforms, challenges like communication latencies, function scheduling, and data access patterns are further increased. Deploying heterogeneous functions on these target platforms can make these challenges even harder to solve~\cite{malawski2017benchmarking}. 
However, the opportunities that the FDN offers such as varied target platform's characteristics, possibility of collaborative execution between multiple target platforms, and localization of data can help in achieving higher Service Level Objective (SLO) matching, lower energy consumption and higher throughput for a mix of applications.

Extending FaaS to the continuum of resources requires scheduling functions and placing data onto the target platforms. This requires more knowledge about the behavior of the application functions. The assumption of similar granularity does not hold since applications will use functions with significantly different computational requirements. Furthermore, data will be stored in different databases at different locations providing a non-uniform access latency. Although the functions are stateless, state changes and look-ups require frequent access to databases. The data access behavior of functions is not taken into account by the current platforms for scheduling~\cite{hellerstein2018serverless}. Therefore, when scheduling function invocations both, the computational and the data requirements have to be considered in an optimized manner benefiting from the distribution and the heterogeneity of the compute and data resources.

\begin{figure}[t]
     \centering
     \includegraphics[width=\linewidth]{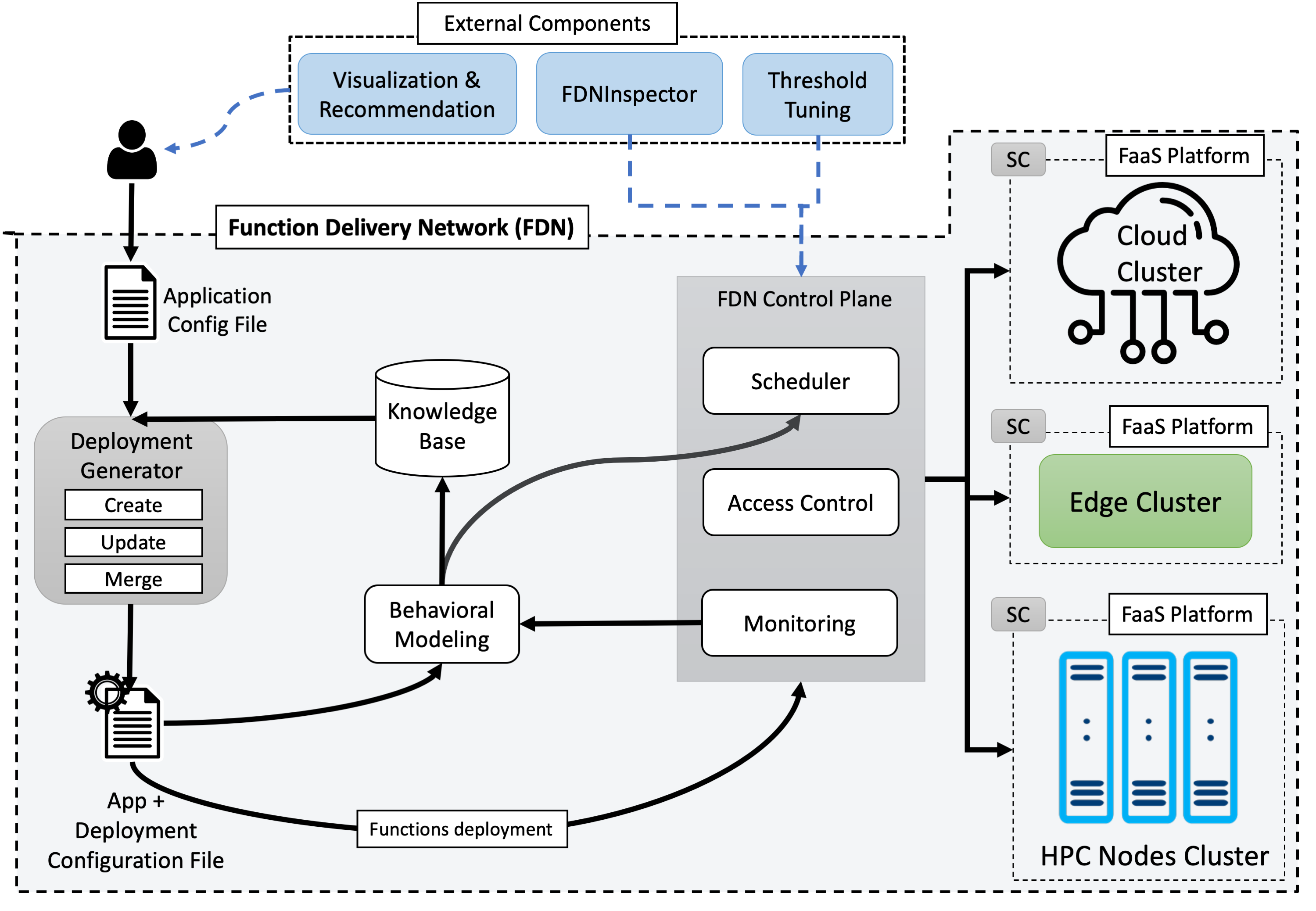}
     \caption{Overall architecture and high level workflow of the Function Delivery Network (FDN). The FDN combines several target platforms (Cloud, Edge, HPC) via a joint Control Plane for the continuous deployment of applications into the Computing Continuum. It analyzes application characteristics (Behavioral Modeling) and the FDN platform parameters (Monitoring) and applies a distributed approach for function scheduling (Scheduler and Sidecars in the platforms). External tools, such as the FDNinspector presented in this paper, allow to benchmark the FDN and tune its hyperparameters.} \label{fig:FDNarchitecture}
\end{figure}

Integrating target platforms with different levels of computing power has the potential to improve overall application performance. Different types of hardware may reduce the overall energy consumption by integrating IoT and other low-power target platforms~\cite{energyfog}. In the same way, high-performance computing target platforms might add large amounts of computing power. Other domains where heterogeneous FaaS platforms can be relevant are edge and fog computing. Both domains include several different types of hardware nodes, sometimes with a huge difference in computing power (e.g. a smartphone and AWS~\cite{edgecomputing}). Situations like these arise the need for intelligent placement of computational tasks to fully exploit the benefits of edge and fog computing. In these domains, network links might vary vastly in terms of bandwidth sizes (e.g. Infiniband in datacenters and mobile networking in cellphones)~\cite{fogcomputing}. This improves the need for efficient scheduling of functions to reduce the overall latency for the participants of the heterogeneous target platforms. Integrating specialized FaaS platforms which operate better on specific hardware (e.g. an HPC cluster or a cluster of IoT devices) can leverage both dimensions of heterogeneity and can optimally exploit the available resources.

Figure~\ref{fig:FDNarchitecture} outlines the overall architecture and high-level workflow of the proposed Function Delivery Network (FDN). The user provides an application \textit{configuration specification} which describes the functions, APIs, permissions, configurations, and events. The specification can be given via a command-line or web interface, or by using some frameworks like Serverless~\cite{serverlessFramework} and Architect~\cite{architectFramework}. The \textit{Deployment Generator} annotates this file with the deployment configuration either based on the previous knowledge captured in the \textit{Knowledge Base} or based on the expert knowledge provided externally. This updated  specification is then passed to the \textit{FDN Control Plane}. It manages function scheduling and data placement, monitors the overall infrastructure and applications, and provides access control for authentication and authorization. The functions are scheduled to the target platforms based on the specification. Various behavioral models are constructed during application execution by the \textit{Behavioral Modeling} component. These models are updated regularly in an online learning manner as the data from the application functions is collected. The runtime decisions of function scheduling and data placement done by the \textit{FDN Control Plane} is based on these models. Furthermore, the gathered historic application knowledge is used by external components for recommendations to the user, or for offline tuning of the FDN itself. The \textit{FDNInspector} (presented in Section~\ref{sec:fdn_inspector}) an external component of the FDN is utilized for benchmarking the FDN . The following subsections describe each component of the FDN in more detail.

% These models are updated regularly in online learning fashion as the application functions data is collected.

% The runtime decisions of functions scheduling and data placement by \textit{FDN Control Plane} is based on these models.  Furthermore, the gathered application historic knowledge is used by the external components for recommendations to the user, or for offline tuning of the FDN itself, and benchmarking the FDN using the \textit{FDNInspector}.

% Following subsections describe each component of the FDN in more detail. 

\subsection{FDN Control Plane}
\label{sec:fdn_control_pane}

% This is the main component of the FDN and is responsible for managing the FDN. Its responsibilities include access control for authentication and authorization, monitoring across the different target platforms, and scheduling function invocations and placement of data. The management of FaaS platforms among the target platforms is mostly situated within the individual FaaS platform, however, a part of the management of FaaS platform is delegated to FDN control plane.  The collaboration between the control plane and the local FaaS platform is performed through a Sidecar Controller Component on each of the target platform.
% The details of each sub-components of FDN control plane is described below. 
This is the main component of the FDN and is responsible for managing the FDN. It's responsibilities include access control for authentication and authorization, monitoring across the different target platforms, and scheduling function invocations and placement of data. The management of target platforms is done in a hierarchical manner, where the scheduling and placement decisions concerning the target platforms are taken by the scheduler within this component, while the selection of the nodes within the target is delegated to the \textit{Sidecar Controller} component within each target platform. Both the control plane and the local sidecar controller work in collaboration to the make final decision. The details regarding each sub-component of the FDN control plane are presented below. 

\subsubsection{Access Control} 
\label{sec:access_control}
Every individual computing platform requires some security measures for scheduling functions and collecting resource utilization data from them. This component deals with these measures. 

\subsubsection{Monitoring} 
\label{sec:monitoring}

% This sub-component is responsible for gathering platform, application, and function level metrics data. For collecting a wide range of metrics, it interfaces and extends the existing monitoring of the FaaS platforms and the Kubernetes clusters and provides base data for the \textit{Scheduler} and the \textit{Behavioral Modeling} components. Prometheus is used along with some added instrumentation for collecting heterogeneous monitoring data. Metrics are classified under three categories: 1) User-Centric metrics, the metrics responsible at the user side, 2) FaaS-Platform-Centric metrics, the metrics from the FaaS platform, and 3) Infrastructure-Centric metrics, the metrics from the host machines. Table~\ref{monitoring_metrics} shows the currently considered metrics from three different categories and these metrics data is collected per unit time. 
This component is responsible for gathering data related to platform, application, and function level metrics. For collecting a wide variety of metrics, it interfaces and extends the existing monitoring of the FaaS platforms and the Kubernetes clusters, and provides base data for the \textit{Scheduler} and the \textit{Behavioral Modeling} components. Prometheus being a well known monitoring system will be used along with some added instrumentation for collecting heterogeneous monitoring data~\cite{turnbull2018monitoring}. However, in this work we have already built a monitoring system based on Prometheus as part of the FDNInspector (Section ~\ref{sec:fdn_inspector}) to extract different metrics for evaluation. This will be reused for the implementation of the FDN. Metrics are classified under three categories: (i) User-Centric metrics, the metrics responsible at the user side, (ii) FaaS-Platform-Centric metrics, the metrics from the FaaS platform, and (iii) Infrastructure-Centric metrics, the metrics from the host machines.  
\begin{itemize}
     \item \textbf{User-Centric metrics}: The response time for a HTTP request below which 90\% of the response time values lie, is called the 90-percentile (P90) response time, which means 90 percent of the requests are processed in 90-percentile response time or less. This metric is important from the SLA point of view, where one wants to have most of the requests (90\% in this case) completed before a certain time. This metric and the number of requests served per unit time are calculated as part of this class of metrics. 
     \item \textbf{Platform-Centric metrics}: Number of function invocations resulted from the received requests, number of replicas for the function creation created to load balance those invocations, number of invocations resulting in cold starts, and execution time of the function (excluding the startup latency) along with the memory allocated to each function instance is considered in this class of metrics.
     
      \item \textbf{Infrastructure-Centric metrics}: In this case, the amount and usage over time of static resources such as number of cores, memory inside individual nodes of a target platform are considered when functions are scheduled on it.
    %   Here the target platform's nodes static resources such as the number of cores, memory is considered along with the usage of those resources over time when the function is scheduled on it. 
\end{itemize}
The summary of these considered metrics from the three different categories are shown in Table~\ref{monitoring_metrics}. For all these metrics, the data is collected per unit time. 
\begin{center}
\begin{table}[t]%
\centering
\caption{Monitoring metrics from three different layers.\label{monitoring_metrics}}%
\begin{threeparttable}
\begin{tabular*}{500pt}{@{\extracolsep\fill}lll@{\extracolsep\fill}}
\toprule
\textbf{User-Centric Metrics} & \textbf{FaaS-Platform-Centric Metrics}  & \textbf{Infrastructure-Centric Metrics}  \\
\midrule
Requests  90-percentile (P90) response time & Number of function replicas  & Total number of cores    \\
Number of requests served  & Number of function invocations  & Total Memory \\
& Number of cold starts   & CPU utilization of cluster   \\
& Function execution time  & Memory utilization of cluster    \\
& Memory allocated to the function &  Disk I/O of cluster  \\
\bottomrule
\end{tabular*}
\end{threeparttable}
\end{table}
\end{center}

% \hlcyan{Furthermore, monitoring data like events (allocation of resources, start of container, deletion of container, etc.) can also be included in it}. These events data is helpful for building models for anomaly detection and finding the root cause analysis. The extensions for monitoring solution for collecting such metrics must be carefully designed to reduce application jitter and performance degradation. All the collected monitoring data is stored inside the database and is used by the FDN's \textit{Behavioural Modeling} (Section ~\ref{sec:behaviour_modelling}) component for building the models. 
Tracing of events (allocation of resources, start of container, deletion of container, etc.) will be added in the future, since these events are helpful for building models for anomaly detection and finding the root cause analysis. The monitoring solution must be carefully designed to reduce application jitter and performance degradation. All the collected monitoring data are stored inside the database and are used by the FDN's \textit{Behavioural Modeling} (Section ~\ref{sec:behaviour_modelling}) component for building various models.

% These events data is helpful for building models for anomaly detection and finding the root cause analysis. The extensions for monitoring solution for collecting such metrics must be carefully designed to reduce application jitter and performance degradation. All the collected monitoring data is stored inside the database and is used by the FDN's \textit{Behavioural Modeling} (Section ~\ref{sec:behaviour_modelling}) component for building the models. 

\subsubsection{Scheduler}
\label{sec:scheduler}
It is responsible for (1) scheduling or delivering the function and (2) placement of the data to an appropriate target platform based on the compute and data requirements of the function. Apart from function scheduling and data placement, this component also keeps track of the high availability of the applications. For taking decisions, this component uses the data from the \textit{Monitoring} (Section ~\ref{sec:monitoring}) and the \textit{Behavioural Modeling} (Section~\ref{sec:behaviour_modelling}) components, and applies a hierarchical decision making approach. In this approach, the scheduling and placement decisions with respect to the target platform are taken by the scheduler, while the selection of the resource within the target platform is delegated to the \textit{Sidecar Controller} component. The three important functionalities of the \textit{scheduler} are described below:  

% where scheduling and placement decisions with respect to the target platform are taken by the scheduler, while it delegates the selection of the resource within the target to the \textit{Sidecar Controller} component. 

% The three important functionalities of it are described below:  

% It is responsible for scheduling the function and placement of the data to an appropriate target platform based on the compute and \hlcyan{storage} resource requirements of the function. \hlcyan{Apart from function scheduling and data placement, this component also keeps track of the high availability of the applications}. For taking decisions, this component uses the data from the \textit{Monitoring} (Section ~\ref{sec:monitoring}) and the \textit{Behavioural Modeling} (Section ~\ref{sec:behaviour_modelling}) components, and applies a hierarchical decision approach where scheduling and placement decisions with respect to the target platform are taken by the scheduler, while it delegates the selection of the resource within the target to the \textit{Sidecar Controller} component. The three important functionalities of it are described below:  

\paragraph{Function Scheduling} 
The \textit{Scheduler} is responsible for scheduling the function to the right target platform based on a distributed scheduling algorithm. This algorithm uses the function's behavioral models, the target platforms configuration and  the current state of the FDN for making a decision.
Additionally, it investigates the trade-offs between staging the data for individual function invocations or long term migration to a specific server within the target platform for faster data access, and then selects the best suitable option.  

\paragraph{Data Placement}
Data Placement functionality includes tools and methods for adaptive data management. It enables migration of data between the target platforms to exploit data affinity. Targets of the adaptive data management are mostly the NoSQL databases and object storage platforms such as MinIO~\cite{MinIO:online}, that are used for storing the state of the functions and data files. It includes following three main methods for data management: 

% Data Placement functionality includes tools and methods for adaptive data management. It enables migration of data between the target platforms to exploit data affinity. Targets of the adaptive data management are mostly the NoSQL databases and object storage platforms, they are used for storing the sate of the functions and data files. It includes following three main methods for data management : 

\begin{enumerate}[1.]
 \item \textbf{Distributed Data Caching}: It is used for supporting data affinity. It acts as an intermediate layer between functions and the used storage platform (databases or object storage). Written data as well as the accessed data will be cached and, means will be provided to proactively migrate or replicate data to the selected target platform for future function invocations.
 %methods
 
%  \item \textbf{File Staging and Migration}: File staging and migration is used where data to be used is stored as files and is accessed by compute intensive functions. Such function invocations are candidates for being scheduled to an HPC target platform. Executing functions on the HPC target platform require staging of input and output files to and from the parallel file system of the cluster. Ideally staging is not done on-demand but proactively according to decisions taken in the resource management. This method is responsible for providing the infrastructure for file staging including a control interface for the resource management.
 \item \textbf{File Staging and Migration}: It is applied when the data to be used is stored in files and is accessed by compute intensive functions. Such function invocations are candidates for being scheduled to a target platform of HPC nodes. Ideally staging is not done on-demand but proactively, and scheduling decisions might even lead to a migration of files for reducing the staging overhead in case of repetitive execution of those functions. 

 \item \textbf{Data Access Instrumentation}: To enable distributed data caching for NoSQL databases as well as file staging, database and file accesses in the functions have to be redirected to the data management layer. This method automatically instruments the deployment specification which will be as transparent as possible for the application developer. The general approach for this method is based on automatically intercepting the REST call and file system functions through library interposition.
\end{enumerate}

Function scheduling and data placement decision methods works in collaboration with each other by taking into account multiple objectives like compute and storage requirements, communication between functions, and cost. 

\paragraph{Fault Tolerance}
Methods for setting up a fault tolerant environment where the failing of a specific device/node in a target platform will lead to a restart or continuation at another device/node in the same or different target platform such that the system continues to operate is done as part of this functionality. It also includes algorithms to detect failures in advance to keep a high availability using the models from \textit{Behavioural Modeling} (Section~\ref{sec:behaviour_modelling}) component.

\subsection{Sidecar Controller (SC)}
\label{sec:sidecar_controller}

% This component resides along with the local FaaS platform installation on the target platform where it acts as a local decision maker. While the control plane takes the decision to which target platform an invocation goes, the local decision to select a node of the Kubernetes cluster is taken locally by this component. Furthermore,it also checks whether to schedule a locally triggered function locally or to delegate to the higher level Control Plane. 

This component resides along with the local FaaS platform installation on the target platform where it acts as a local decision maker. While the \textit{Control Plane} is responsible for deciding the target platform on which the function invocation goes to, the local decision to select a node of the Kubernetes cluster is taken by this component. Furthermore, it also checks whether to schedule a locally triggered function locally or to delegate it to the higher level \textit{Control Plane}. 

% While the control plane takes the decision to which target platform an invocation goes, the local decision to select a node of the Kubernetes cluster is taken locally by this component. 

% Furthermore,it also checks whether to schedule a locally triggered function locally or to delegate to the higher level Control Plane. 

\subsection{Behavioral Modeling}
\label{sec:behaviour_modelling}

This component is responsible for characterizing the behaviour of the function based on the monitoring information (from the  \textit{Monitoring} component) and the deployment configuration file. It characterizes the application by the following models:
% It will characterize the application by the following models: 
\begin{enumerate}[1.]
 \item \textbf{Application Event Model}:  Information about events like frequency of function invocations, sequence of functions invoked, creation, deletion or upgrade of functions is used to build this Model. This model will then be used for use cases like anomaly detection, forecasting of future events, events tracing etc. This event model will also be used for reducing the cold start time by predicting the workload and starting the function containers before time.
 
 \item \textbf{Function Interaction Model}:  It characterizes the producer-consumer interactions of functions based on data accesses. The interaction might, for example, suggest to package functions together to reduce the communication costs.
 
  \item \textbf{Data Access Model}:   It characterizes the functions with respect to their data accesses.  It determines, for example, how frequently data is read or written to certain databases or files. This can be useful for placement of functions considering the caching scenarios.
  
    % \item \textbf{Function Performance Model}:    The Function Performance Model will capture the performance with respect to time and energy for certain combinations of resources, such as the number of cores, the network bandwidth, the memory size and I/O bandwidth. The model will be based on measured information from the FDN monitoring as well as on prediction for other combinations using machine-learning. This model will be used by the scheduler to find the right machine for invoking the function (called as function placement in this context) based on the resource requirement and the availability.
    
    \item \textbf{Function Performance Model}: The Function Performance Model will capture the performance with respect to time and energy for certain combinations of resources, such as the number of cores, the network bandwidth, the memory size and I/O bandwidth. The model will be based on measured information obtained from the FDN \textit{Monitoring} (Section~\ref{sec:monitoring}) as well as on the current workload. This model will be used by the \textit{Scheduler} (Section~\ref{sec:scheduler}) to find the right target platform for invoking the function (called as function delivery in this work) based on resource requirements and availability.
    
    % The model will be based on measured information from the FDN monitoring as well as on prediction for other combinations using machine-learning. This model will be used by the scheduler to find the right machine for invoking the function (called as function placement in this context) based on the resource requirement and the availability.
\end{enumerate}
The models will be provided to the \textit{Scheduler} (Section~\ref{sec:scheduler}) and will be stored in the \textit{Knowledge Base} (Section~\ref{sec:knowledge_base}).  

\subsection{Knowledge Base}
\label{sec:knowledge_base}

% This component stores the application models as well as decisions taken by the scheduler. It automatically adds information to the deployment configuration in case of redeployments or updation of the deployments. Furthermore, the external components, use this information for further analysis and decision making. Scalable NoSQL and SQL databases along with the scalable file storage platforms are the basis for the implementation of the knowledge base.

This component stores the application models prepared by \textit{Behavioral Modeling} (Section~\ref{sec:behaviour_modelling}) as well as decisions taken by the \textit{Scheduler} (Section~\ref{sec:scheduler}). The previously saved high performant decisions are used by the \textit{Deployment Generator} (Section~\ref{sec:dep_gen}) for automatically adding annotations to the deployment configuration in case of redeployments or on deployment updates of the application. Furthermore, the external components, use the stored information for further analysis and decision making. Scalable NoSQL and SQL databases along with the scalable file storage platforms are the basis for the implementation of the \textit{Knowledge Base}.

% Furthermore, the external components, use this information for further analysis and decision making. Scalable NoSQL and SQL databases along with the scalable file storage platforms are the basis for the implementation of the knowledge base.

\subsection{Deployment Generator}
\label{sec:dep_gen}

% This component is responsible for annotating the deployment specification provided by the user. The provided application configuration file can describe the initial application but also serves to specify updates to the already running applications. The Deployment Generator Component inserts hints into the deployment specification like where to deploy a function as well as function and data characteristics. It therefore add annotations based on analysis results in the knowledge base for previous deployments. This is especially important for update deployments modifying the running application. The deployment generator also performs any required instrumentation, for example, to enable data caching and migration. 

The \textit{Deployment Generator} is responsible for annotating the deployment specification provided by the user. The provided application configuration file can describe the initial application deployment configuration but also serves as a measure to specify updates to the already running applications. The \textit{Deployment Generator} component inserts hints into the deployment specification like where to deploy a function as well as function and data characteristics. Therefore, it adds annotations based on analyzing the results in the \textit{Knowledge Base} for previous deployments. This is especially important for deployment updates that modify the running application. It also performs any required instrumentation to the application, for example, to enable data caching and migration. 

\subsection{External Components}
\label{sec:exten_components}

% There are some external components which are either aid to FDN for better decision making or to the user by recommending some deployment configurations for optimization of the function deployment, explanation of the decision by some visualizations or to do the benchmarking of the overall FDN with various applications and functions. Following sub-components are part of it:  
There are some external components which either aid the FDN to take better decisions or help the user by recommending some deployment configurations for optimizing function deployment, explaining runtime decisions through some visualizations, or benchmarking the overall FDN with various applications and functions. Following components are part of it: 

%This para is weird....

% explanation of the decision by some visualizations or to do the benchmarking of the overall FDN with various applications and functions. Following sub-components are part of it: 

\begin{itemize}
    \item \textbf{Recommendation and Visualization}: It extracts data from the \textit{Knowledge Base}, explains the FDN runtime decisions to the user, and recommends certain configurations optimizing the application deployment. Such visualizations can be helpful to the application developer for knowing where the functions are scheduled and based on that some optimizations related to the system architecture can be added.
    
    \item \textbf{FDNInspector (Benchmarking)}: This external component is responsible for  benchmarking the FDN on certain functions and applications.  The benchmarking results can be further used for comparing the application performance by the user on various target platforms. In this work, this component is described in more detail in Section~\ref{sec:fdn_inspector}, where we utilize it, to show the opportunities offered by the FDN in achieving different objectives.
    
    % \item \textbf{Threshold Tuning}: The decisions taken by the scheduler are based on the thresholds deciding, when to, for example, migrate data to a different compute platform. A tuning based on historic data of the FDN will improve the effectiveness of resource management across different applications. This tuning is part of this component. 
    \item \textbf{Threshold Tuning}: The decisions taken by the \textit{Scheduler} (Section~\ref{sec:scheduler}) are frequently based on thresholds that decide, when to, for example, migrate data to a different target platform. A tuning based on historic data of the FDN will improve the effectiveness of resource management across different applications. This tuning is part of this external component.
    \end{itemize}

\section{Methodology}
\label{sec:methodology}

\begin{center}
\begin{table}[t]%
\centering
\caption{ List of FaaS based functions we
developed or modified for demonstrating the opportunities offered by the FDN. \label{functions_used}}%
\begin{threeparttable}
\begin{tabular*}{500pt}{@{\extracolsep\fill}llc@{\extracolsep\fill}}
\toprule
\textbf{Function Name} & \textbf{Description}  & \textbf{Language runtime}\\
\midrule
 nodeinfo   & Gives basic characteristics of node like CPU count, architecture, uptime. & Node.js   \\
primes-python  & Calculates prime numbers till 10000000. & Python3   \\
\multirow{2}{*}{image-processing} & Reads an image from object storage (here Minio) and performs basic   operations & \multirow{2}{*}{Python3}\\
& (flip, rotate, filter, grayscale and resize) to the image. &   \\
sentiment-analysis & Sentiment analysis of the given text.  & Python3    \\
\multirow{2}{*}{JSON-loads} & Takes a big JSON file as input containing 1000 three coordinates (x,y,z)  & \multirow{2}{*}{Python3}   \\
& records and return the average coordinates value. &   \\
 
\bottomrule
\end{tabular*}
\end{threeparttable}
\end{table}
\end{center}

In this section, we first present details about the different benchmarks used for evaluating various opportunities offered by the FDN and then describe the five different target platforms used in this work. We also present the load testing details used for the evaluation, and finally the high-level design and the functioning of the developed \textit{FDNInspector}.

\subsection{Benchmarks}
\label{sec:function_benchmarks}

To investigate the performance of each target platform available as part of the FDN, we used a subset of the benchmarks provided with the FaaSProfiler~\cite{shahrad2019architectural} and modified them for our use case. Furthermore, we developed OpenFaaS implementations of the chosen functions to enable their execution on the target platforms using the OpenFaaS platform. The OpenWhisk action container generally includes code for the function along with its language runtime. OpenWhisk processes the incoming HTTP requests for the function invocation with any number of arguments and sends the results back to the user or caller. For most of the functions, we have used the default runtime environment provided by OpenWhisk depending on the language that the function is written in. If a function uses some extra packages which are not part of their default language runtimes, we created a docker runtime for it based on their default docker runtime. The OpenFaaS functions are also similar to OpenWhisk, however, we created our own docker images for functions, to run them on the ARM  platform. The Google Cloud Functions are also similar to OpenWhisk functions, however, one cannot create their own docker image of the runtime. 
The functions used as part of this work are summarized in the Table~\ref{functions_used} along with their description and language runtimes. The \texttt{nodeinfo} function exposes an HTTP endpoint and provides the user with basic information about the system such as hostname, underlying architecture, number of CPUs, etc. We utilize this function to test the general performance of each target platform and get an overall idea of their capabilities. The  compute-intensive \texttt{primes-python}, \texttt{sentiment-analysis} and \texttt{JSON-loads} functions are used for comparing the high-end target platforms (without the edge-based target platforms). Finally, for demonstrating the advantage of data localisation opportunity in the FDN the object (in our work an image) access latency from the MinIo~\cite{MinIO:online} object storage platform, is showcased using the \textit{image-processing} function.

\subsection{Experimental Target Platforms}
\label{sec:experi_sys_overview}

To demonstrate the opportunities offered by the FDN, we evaluate the function benchmarks on five different target platforms ranging from a high performance HPC node to resource-constrained edge devices. The configuration of each target platform, type of FaaS platform used, and the number of nodes present in that target platform is shown in Table~\ref{clusters_evaluated}. 

The \textit{edge-cluster} consists of three embedded Nvidia Jetson Nano devices~\cite{jetsonano}. Due to the limited resources available on these boards, it was not possible to run the heavy OpenWhisk platform on our \textit{edge-cluster}, and OpenFaaS does support low-end devices and provides binaries for ARM processors, therefore we utilized OpenFaaS on top of k3s~\cite{rancherk24:online}, a light weight version of Kubernetes to host a Kubernetes cluster on it.   k3s reduces the footprint and bootstrap-process of Kubernetes and combines all the low-level components required for running a Kubernetes cluster such as \textit{containerd}, \textit{runc}, and \textit{kubectl} into a single binary.

The \textit{cloud-cluster}  is composed of three virtual machines hosted on a private cloud at the Leibniz Supercomputing Center (LRZ)~\cite{LRZLeibn51:online}. Each VM has four virtual CPU cores and 8 GiB of memory. This target platform is based on OpenWhisk on top of Kubernetes. Additionally, we used Google Cloud Functions (GCF) for creating  \textit{google-cloud-cluster} platform and internal configuration details of the VMs or the containers in which the functions are deployed is not available for the user. 

The other two  target platforms represent compute nodes from High Performance Computing (HPC) environments. The \textit{hpc-node-cluster} is a dual-socket system, with each socket containing an Intel Cascade Lake processor with 22 cores and the \textit{old-hpc-node-cluster} consists of four sockets, with each socket containing an Intel Westmere-EX processor with 10 cores. We disabled hyper-threading and turbo boost on both the HPC clusters. OpenWhisk on top of Kubernetes is deployed on each of these nodes.

\begin{center}
\begin{table}[t]%
\centering
\caption{Different target platforms used as part of this work for evaluating the benchmarks. \label{clusters_evaluated}}%
\begin{threeparttable}
\begin{tabular*}{500pt}{@{\extracolsep\fill}llllc@{\extracolsep\fill}}
\toprule
\textbf{Target Platform} & \textbf{Processor}  & \textbf{FaaS Platform} & \textbf{H/W Specifications} & \textbf{Nodes in Cluster}\\
\midrule
\multirow{2}{*}{hpc-node-cluster} & Intel(R) Xeon(R)   & \multirow{2}{*}{OpenWhisk} & 44 Cores  & \multirow{2}{*}{1}   \\
 & Gold 6238 CPU @ 2.10GHz &  & 754 GiB memory &    \\
 
 \multirow{2}{*}{old-hpc-node-cluster} & Intel(R) Xeon(R)  & \multirow{2}{*}{OpenWhisk} & 40 Cores  & \multirow{2}{*}{1}   \\
 & CPU E7- 4850  @ 2.00GHz &  & 251 GiB memory &    \\
 
  \multirow{2}{*}{cloud-cluster} & Intel(R) Xeon(R)  & \multirow{2}{*}{OpenWhisk} & 4 vCPU  & \multirow{2}{*}{3}   \\
 & CPU E5-2697A v4 @ 2.60GHz  &  & 8 GiB memory &    \\
 
   \multirow{1}{*}{google-cloud-cluster\tnote{$\dagger$}} & N/A\tnote{$\dagger$}  & \multirow{1}{*}{GCF} & N/A\tnote{$\dagger$}  & \multirow{1}{*}{N/A\tnote{$\dagger$}}   \\
 
  \multirow{2}{*}{edge-cluster} & ARMv8 Processor  & \multirow{2}{*}{OpenFaaS} & 4 Cores  & \multirow{2}{*}{3}   \\
 & rev 1 (v8l)  &  & 4GiB memory &    \\
 
\bottomrule
\end{tabular*}
\begin{tablenotes}
\item[$\dagger$] Host VMs or containers configuration information in which functions are deployed is not available. 
\end{tablenotes}
\end{threeparttable}
\end{table}
\end{center}
\subsection{Load Test Settings}
\label{sec:load_test_settings}

The evaluation of various opportunities was done using the free and open-source load testing tool, \textit{k6}~\cite{k6:online}. \textit{k6} uses a script for running the tests where the HTTP(s) endpoint along with the request parameters are specified. HTTP(s) endpoint represents the deployed function endpoint and varies with each function and target platform in our work. Two of the other k6 parameters which are configured as part of each test are: 
\begin{itemize}
    \item \textbf{Virtual Users (VUs)}: Virtual Users (VUs) are the entities in k6 that execute the test and make HTTP(s) or websocket requests. VUs are concurrent and will continuously iterate through the request endpoint until the test ends.

    \item \textbf{Duration}: A string specifying the total duration a test will run. During this time each VU will execute the script in a loop.
\end{itemize}

In our evaluations, duration was fixed to \texttt{10 minutes} and number of VUs varied from \texttt{10 to 50} depending on the function and the target platform. The total duration for which the metrics data is collected is set to \texttt{20 minutes} and the sampling rate is set to \texttt{10 seconds}, i.e, metrics values are aggregated for 10 seconds. The term unit time refers to the sampling interval in Section~\ref{sec:results}.

The number of requests per second generated by k6 depends on the number of VUs and the time taken by each request to complete.
For example, if there are 10 VUs with total test duration set to 10 minutes and each request from a VU took 30 seconds to complete, then from each VU there will be 2 requests per minute and 20 requests per minute from 10 VUs with a total of roughly 200 requests completed in the whole duration. Therefore it will vary for each target platform depending on the time taken by each request to complete.

Moreover, we increased the default limits on the number of concurrent invocations and invocations per minute which can be served in OpenWhisk to \texttt{99999} and increased the memory allocated to the invoker to \texttt{4096} MiB for each target platform using OpenWhisk.

% List of FaaS based functions we
% developed or modified for showcasing the requirement for FDN.

% \begin{center}
% \begin{table}[t]%
% \centering
% \caption{Different target platforms used as part of this work. \label{clusters_evaluated}}%
% \begin{threeparttable}
% \begin{tabular*}{500pt}{@{\extracolsep\fill}llclc@{\extracolsep\fill}}
% \toprule
% \textbf{Target Platform} & \textbf{Processor}  & \textbf{FaaS Platform} & \textbf{H/W Specifications} & \textbf{Nodes in Cluster}\\
% \midrule
% \multirow{2}{*}{hpc-node-cluster} & Intel(R) Xeon(R)   & \multirow{2}{*}{OpenWhisk} & 44 Cores  & \multirow{2}{*}{1}   \\
%  & Gold 6238 CPU @ 2.10GHz &  & 754GiB memory &    \\
 
%  \multirow{2}{*}{old-hpc-node-cluster} & Intel(R) Xeon(R)  & \multirow{2}{*}{OpenWhisk} & 40 Cores  & \multirow{2}{*}{1}   \\
%  & CPU E7- 4850  @ 2.00GHz &  & 251GiB memory &    \\
 
%   \multirow{2}{*}{cloud-cluster} & Intel(R) Xeon(R)  & \multirow{2}{*}{OpenWhisk} & 4 vCPU  & \multirow{2}{*}{3}   \\
%  & CPU E5-2697A v4 @ 2.60GHz  &  & 8GiB memory &    \\
 
%   \multirow{2}{*}{edge-cluster} & ARMv8 Processor  & \multirow{2}{*}{OpenFaaS} & 4 Cores  & \multirow{2}{*}{3}   \\
%  & rev 1 (v8l)  &  & 4GiB memory &    \\
 
% \bottomrule
% \end{tabular*}
% \end{threeparttable}
% \end{table}
% \end{center}

% \hlcyan{Anshul: add here info about load test and explaination about metrics (unit time, VUs, P90 response time and others}
% \hlcyan{Need to explain unit time and other metrics}

\begin{figure}[t]
\centerline{\includegraphics[width=\linewidth]{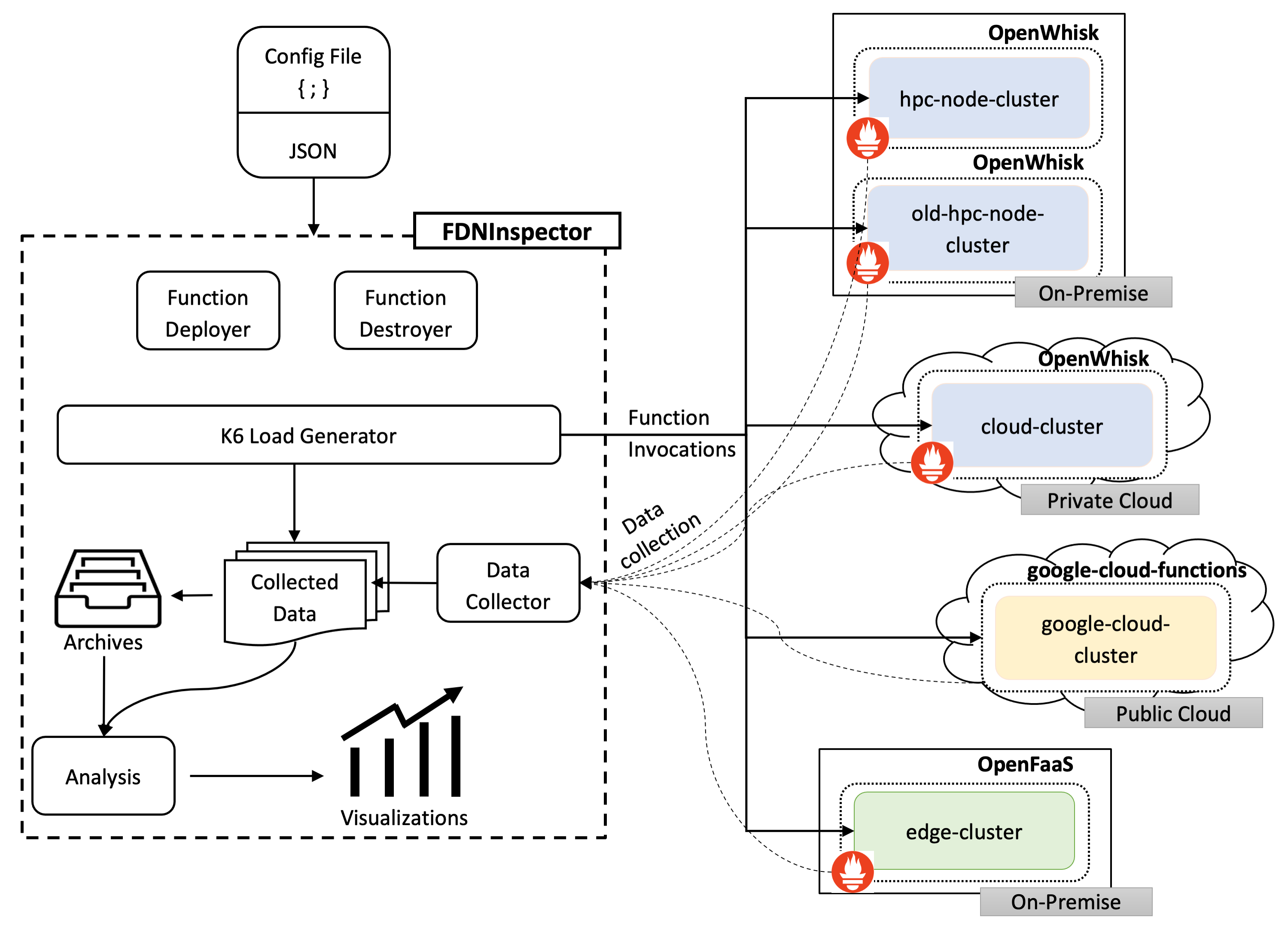}}
\caption{Overall architecture of \textit{FDNInspector} along with the interaction between its components when a load of function invocations is generated for five different target platforms. 
\label{fig_fdn_archi}}
\end{figure}

\subsection{FDNInspector}
\label{sec:fdn_inspector}

\begin{lstlisting}[float, language=json,  caption={Example configuration file in JSON format.}, label=listing:config_file]
{ "test_name": "example_test_name",
  "functions_config: "functions/config.json",
  "target_platforms_config": "target_platforms/config.json",
  "influxdb_url": "http://localhost:8086/",
  "openwhisk_target_platforms": ["cluster-1","cluster-2"],
  "openfaas_target_platforms": ["cluster-3"],
  "public_cloud_target_platforms": ["google_cloud_cluster"],
  "test_instances": {
      "instance1": {
        "application": "primes-python",
        "test_settings": {
          "vus": "30",
          "duration": "600s",
          "param_file": "",
          "sleep": "1"
        }
      }
    }
}
\end{lstlisting}

In this work, we introduce the \textit{FDNInspector}\footnote{https://github.com/ansjin/hetrogenous-faas-profiler}, a tool for benchmarking the different target platforms of the FDN to identify opportunities from smart function scheduling and data placement across the heterogeneous target platforms. This external component of the FDN is built first to support our proposed work towards combing heterogeneous target platforms into the FDN.

\textit{FDNInspector} is written in python and serves the following purposes:  
\begin{itemize}
    \item Facilitating the benchmarking via a centralized tool for remotely deploying and executing FaaS benchmarks on any of the target platforms. 
    % \item Supports multiple FaaS based platforms (OpenWhisk and OpenFaaS) deployment, where other platforms as well can be added in an easy fashion. 
    % \item  A tool for generating a high amount of function invocations for the desired duration and amount on different target platforms and collect a diverse range of metrics data (presented in section), which later on is analyzed and visualized automatically.  
    \item  Enabling load generation through function invocations for the desired duration and amount on different target platforms.
    \item Automatic collection of a diverse range of metrics. 
    \item Visualization of the gathered information for manual analysis.
    \item Supports multiple testing opportunities which can be achieved using the FDN where new ones can be added easily.
\end{itemize}

% Figure~\ref{fig_fdn_archi} shows the overall architecture of teh FDNInspector along with high-level flow of its components interaction with different target platforms when the functions are invoked. In our testbed, \textit{hpc-node-cluster} and \textit{old-hpc-node-cluster} target platforms are running on-premise in the private network with using a proxy server for accessing the internet. \textit{cloud-cluster} is run on the private cloud with 3 VMs and can access the internet directly. Lastly, the \textit{edge-cluster} is also running on-premise and use proxy server for accessing the internet but in a different network then the hpc-node based clusters. FDNInspector was deployed on a machine having the access to all these target platforms. Each of these target platforms is running Prometheus instance for collection of data of various metrics.
Figure~\ref{fig_fdn_archi} shows the overall architecture of the \textit{FDNInspector} and the interaction of its components with different target platforms when functions are invoked.  In our experimental setup, the \textit{hpc-node-cluster} and \textit{old-hpc-node-cluster} target platforms are running on-premise in a private network and use a proxy server for accessing the internet. The \textit{cloud-cluster} is running on the private cloud at LRZ~\cite{LRZLeibn51:online} and can access the internet directly. The \textit{google-cloud-cluster} is running on Google Compute Platform (GCP) in \texttt{us-east} region. Similar to the HPC clusters, the \textit{edge-cluster} is also running on-premise and uses a proxy server for accessing the internet. The \textit{FDNInspector} was deployed on a virtual machine having the access to all these target platforms on-premise (in Germany). Each 
target platforms is running a Prometheus instance to collect data for a variety of  metrics.

The user provides the input file in JSON format, an example of which is in shown in Listing~\ref{listing:config_file}. The  target platform information like host address, authentication, and hardware resources is present in the clusters configuration file (Line 2). Information related to functions such as name, docker image, and runtime is present in the functions configuration file (Line 3). The user provides the function name and the target platforms (based on OpenWhisk, OpenFaaS and public FaaS platform) on which the test is to be executed (Line 5-7). Furthermore, the user provides the test parameters: number of virtual users (VUs)  analogous to the actual users, duration of the test, parameter file (if it exists) and how much time (in seconds) to sleep in-between the requests (Line 8-14). This is particularly useful in cases where requests take a longer time to complete and we do not want to send the next request until certain time has already been passed.

The \textit{Function Deployer} takes this configuration file as input and deploys the functions on the listed target platforms. The \textit{wsk} command-line interface is used for deploying the functions onto the OpenWhisk based target platforms. For deploying functions on the OpenFaaS target platforms, \textit{faas-cli} is used. For deploying functions on the google cloud target platform, \textit{gcloud} is used. Once the functions are deployed, we utilize \textit{k6} for invoking the functions on each of the target platforms based on the input parameters (VUS, duration and sleep\_time). After the load generation is finished, the \textit{Data Collector} collects data for a variety of metrics by querying the Prometheus instance of each target platform. The collected data is  presented to the user through graphs. After completion of the tests, the \textit{Function Destroyer} deletes the function instances from each target platform. 

% and \textit{faas-cli} is used for deployment of functions on target platforms based on OpenFaaS. Once the functions are deployed, \textbf{K6} load generator, a Golang based load generator tool is started to invoke the functions on the target platforms based on the input parameters(VUS, duration and sleep\_time). Once the load generation is completed, the metrics data is collected by the \textbf{Data Collector} by querying the Prometheus instance of each target platform and is then used for analysis either in comparison to the previous results or within itself. This analyzed data is visualized and presented to the user. After completion of the tests, \textbf{Function Destroyer} deletes the function instances from each of the target platforms. 

\section{Results}
\label{sec:results}

\begin{figure}[t]
\centerline{\includegraphics[width=\linewidth]{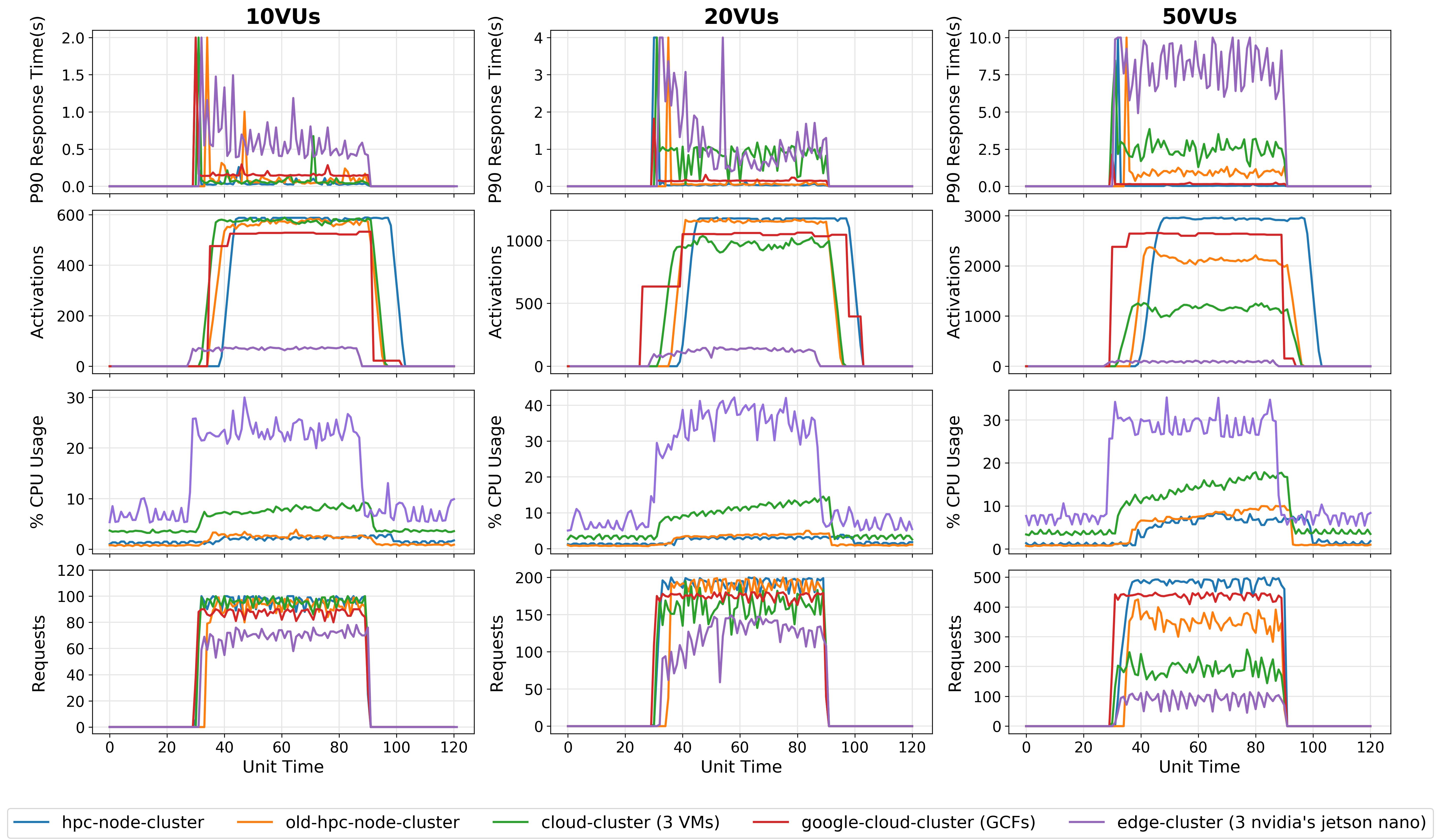}}
\caption{
Comparison between different target platforms on four different metrics (represented as rows) when \texttt{nodeinfo} function is invoked with varied workload (represented as columns). The \textit{edge-cluster} being a resource-constrained target platform exhibits the worst performance in terms of the number of requests processed (maximum 150 requests/second) and their P$90^{th}$ response time (minimum 1s)  among all the five target platforms. All the other target platforms perform similar at a workload less than 50 VUs. However, at 50 VUs workload, the different compute power of each target platform is more prominent (\textit{hpc-node-cluster} and  \textit{google-cloud-cluster}  exhibiting the best performance).  It is to be noted that the $90^{th}$ percentile response time of the requests is truncated above 2, 4 and 10 seconds in each of the workloads to show better comparison after the initial longer requests time due to cold-start.   
\label{nodeinfo_clusters_computational_power}}
\end{figure}

We evaluate the five target platforms for demonstrating the opportunities offered by the FDN for handling function scheduling and data placement in these heterogeneous target platforms in achieving different objectives. However, before evaluating them it is important to know the capability of each target platform. To this end, we evaluate the resource usage of the \texttt{nodeinfo} function on each cluster by varying the number of virtual users (from 10 to 50). Figure~\ref{nodeinfo_clusters_computational_power} shows the result of the conducted evaluation for  different metrics represented as rows (from bottom to top): Number of requests processed, percentage CPU utilization of the cluster, number of activations and the $90^{th}$ percentile (P90) of the response time in seconds of the requests. Different numbers of VUs are represented as columns (increasing from 10 VUs to 50 VUs). The \textit{edge-cluster} exhibits the worst performance in terms of the number of requests processed (approximately 70-150 requests/second) and P90 response time (approximately 1 second and above) across all scenarios. This can be attributed to the limited number of resources and low compute capability of ARM processors as compared to other architectures in our platform. When the number of VUs are less than or equal to 20, the four target platforms: \textit{hpc-node-cluster, old-hpc-node-cluster, google-cloud-cluster and cloud-cluster} perform similar. However, when the load is increased to 50 VUs (last column in  Figure~\ref{nodeinfo_clusters_computational_power}), the different compute capabilities of each target platform is more prominent. The \textit{hpc-node-cluster} performs best and can handle around 500 requests per second with the P90 response time being below 500 ms. The \textit{google-cloud-cluster} performs second-best followed by the \textit{old-hpc-node-cluster} and then the \textit{cloud-cluster} with requests per second and P90 response times being 450, 500ms, 400, 1s, and 200, 2.5s respectively. 

It is apparent from the metric \textit{P90 Response Time}, that the requests initially suffer from the cold-start problem~\cite{mcgrath2017serverless} in all tests for all the target platforms. The initial P90 response time is above five seconds, but after the containers are warm it decreases significantly for all the target platforms. \textit{Activations} represent the number of functions invoked overtime. All the requests in OpenWhisk were sent with the \textbf{blocking} parameter enabled. This means that a function invocation request will wait for the activation result to be available. For OpenFaaS there is no such parameter and it is by default blocking. For the workload with 50 VUs, \textit{hpc-node-cluster} has the highest number of activation's since it serves more number of requests over time as compared to the other clusters. The overall CPU utilization of both HPC node clusters is similar across all tests and is lower than the utilization of the other two clusters. The CPU utilization metric indicates amount of workload a cluster can handle.

\begin{figure}[t]
\centerline{\includegraphics[width=\linewidth]{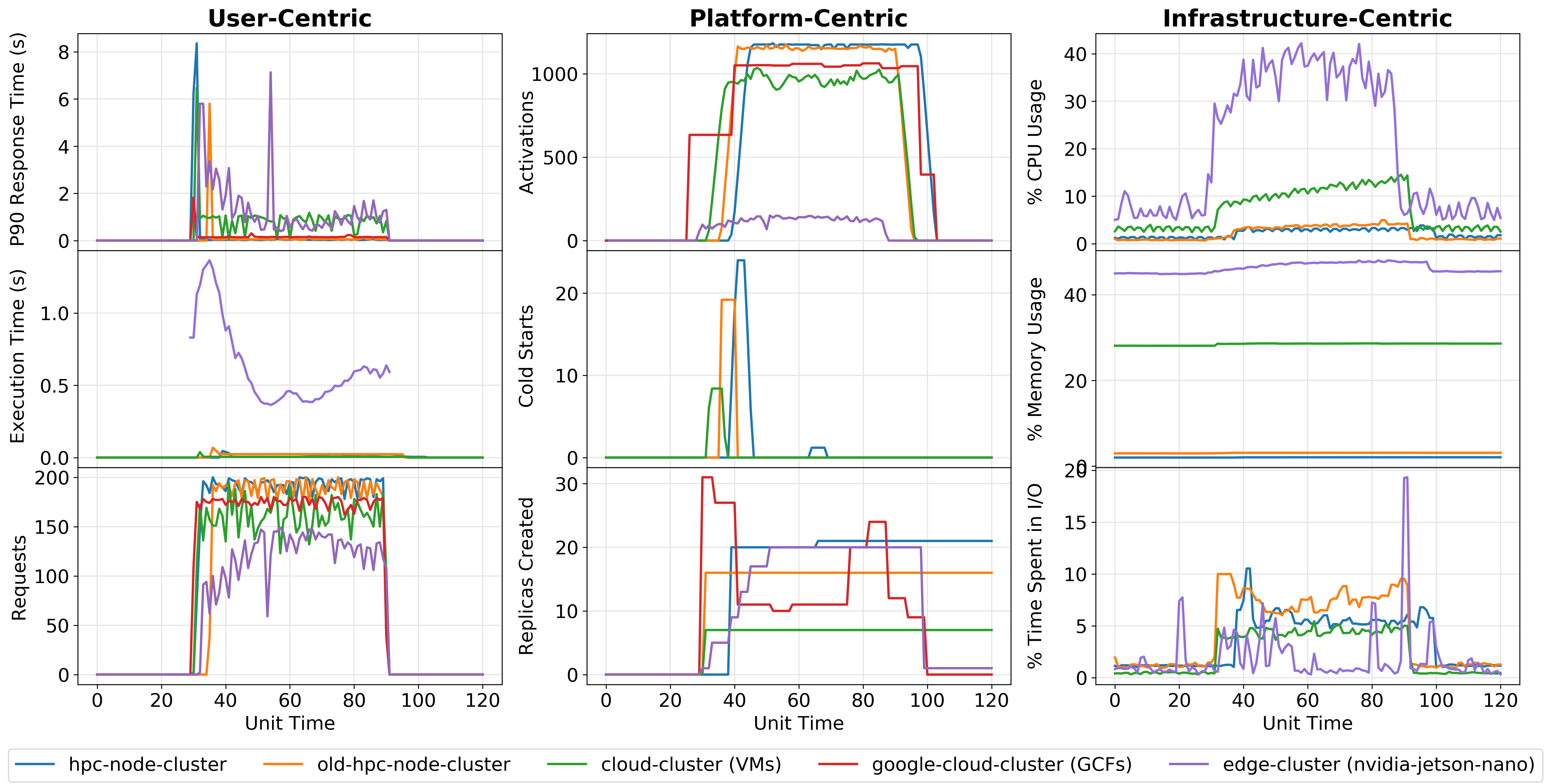}}
\caption{ Comparison of \texttt{nodeinfo} function for all target platforms with workload from 20 VUs on three different classes of the metrics. 
\label{all_metrics_nodeinfo_20VUs}}
\end{figure}

% Figure~\ref{all_metrics_nodeinfo_20VUs} shows the detailed view of all the 9 metrics (divided into 3 classes: user-centric, platform-centric and infrastructure-centric parameters)  when the workload of 20 VUs is applied on the \texttt{nodeinfo} function for all the 4 target platforms. \textit{Execution time} represents the function execution time and \textit{P90 response time} is the difference between the time when the request was sent and it is returned. For each target platform, initial function execution time is high in comparison to the later part of the function invocations due to the initial function cold-starts (2nd row, 2nd column) and the replicas for function were not fully scaled (3rd row, 2nd column). \textit{edge-cluster} being deployed on the OpenFaaS platform, therefore the cold-start metric was not available.  In the infrastructure-centric metrics, there is not much effect for \texttt{nodeinfo} function on the  target platform's memory usage, where as a significant change can be observed in the the CPU and disk I/O usage from their base usages.  Based on these metrics, inter metrics classes relation can be derived like the relation between the number of requests and the replicas created or a relation between the function execution time and the CPU usage of the target platform. These relations can then be used for scheduling the functions to the target platform based on their requirements. In the later half of this section, a subset of these metrics will be used for verifying the need of the FDN under different hypothesis.  

Figure~\ref{all_metrics_nodeinfo_20VUs} shows the detailed view of all the 9 metrics (divided into 3 classes: user-centric, platform-centric and infrastructure-centric parameters) when the workload of 20 VUs is applied on the \texttt{nodeinfo} function for all the five target platforms. \textit{Execution time} represents the function execution time and \textit{response time} is the difference between the time when the request was sent and when it was returned. For each target platform, initially the function execution time is high due to cold-starts (2nd row, 2nd column), and slow increase in the number of replicas (3rd row, 2nd column). Since, the \textit{edge-cluster} was deployed on the OpenFaaS platform, values for the cold start metric were not available due to them being not exposed by OpenFaaS. It is only possible to obtain them through external instrumentation.  Also, various infrastructure level metrics from within the \textit{google-cloud-cluster} requires external instrumental and were not done within the scope of this work. While the \texttt{nodeinfo} function does not have much effect on the memory usage of a target platform, a significant change can be observed in the CPU and disk I/O usage for all target platforms. Based on the values of these metrics, different derived metrics can be formulated. For instance, the relation between the number of requests and the replicas created or a relation between the function execution time and the CPU usage of the target platform.  These derived metrics can then be used for scheduling the functions to the target platform based on their requirements. In the later half of this section, a subset of these metrics is used for demonstrating the opportunities FDN offers in achieving different objectives.  

\begin{figure}[t]
\centerline{\includegraphics[width=\linewidth]{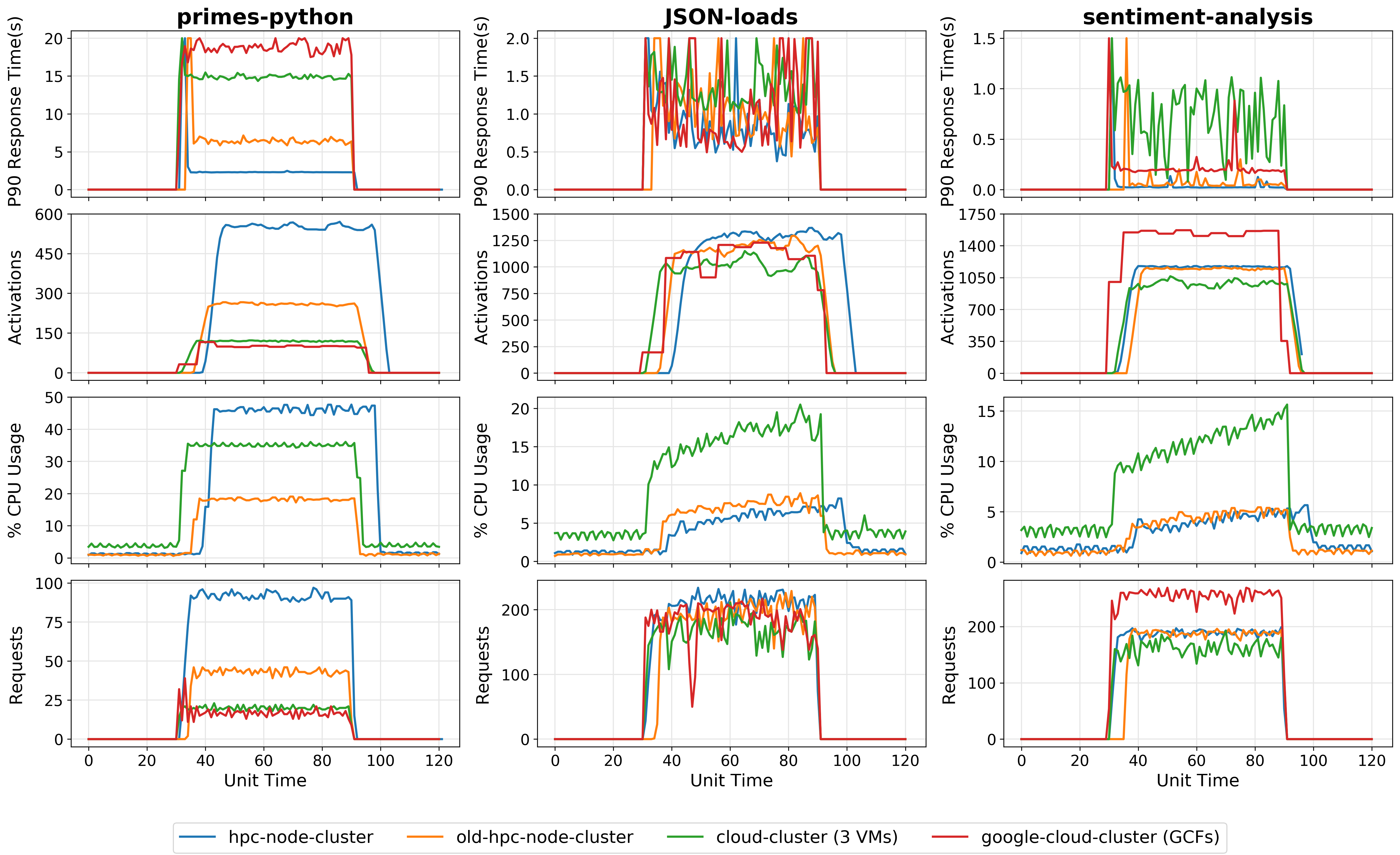}}
\caption{ Comparison of three different functions: \texttt{primes-python}, \texttt{sentiment-analysis} and \texttt{JSON-loads} for three target platforms with 30 VUs generating the load on four different metrics. 
\label{functions_compare_three_target_platforms}}
\end{figure}

\texttt{nodeinfo} being a simple HTTP endpoint function, does not characterize the performance of the target platforms for more complex functions. Towards this, we perform a workload test with 30 VUs using three different functions: \texttt{primes-python and sentiment-analysis} being compute-intensive and \texttt{JSON-loads} being I/O-intensive (see Table~\ref{functions_used}). Figure~\ref{functions_compare_three_target_platforms} shows the results of our experiment, where three different functions are represented as columns and four different metrics are represented as rows. The \textit{edge-cluster} cannot handle a high load for these three different functions, therefore this comparison is only conducted for the four target platforms: \textit{hpc-node-cluster, old-hpc-node-cluster, cloud-cluster, google-cloud-cluster}. The function \texttt{primes-python} is the most compute-intensive with a P90 response time of 14 seconds and 2 seconds per request for the \textit{cloud-cluster} and \textit{hpc-node-cluster} respectively. Also, \textit{google-cloud-cluster} performed worst for this function with 20 requests per unit time and 19 seconds as P90 response time. This could be attributed towards the inability of the GCF to handle compute intensive functions. Furthermore,  this evaluation results demonstrates the higher computation power of the \textit{hpc-node-cluster} as compared to the other target platforms. For the other two functions, all target platforms perform similar to each other with \textit{google-cloud-cluster} performing the best. However, the CPU utilization of \textit{cloud-cluster} is much higher for the \texttt{JSON-loads} and \texttt{sentiment-analysis} functions. Due to high computations requirement for  \texttt{primes-python} function, each target platform is able to process smaller number of requests per unit time (maximum around 100 requests per unit time) as compared to the other functions (maximum around 250 requests per unit time). Such an analysis can be used to derive inter-target platform relations for the same function. Inter-target platform relations can be used to offload a function from one target platform to another based on the function's performance within one target platform and then using these relations to find which target platform will be ideal for it.

% Following subsections presents different goals which can be achieved using the FDN and we evaluated them by using FDNInspector on the 4 target platforms with different functions. 
In the following subsections, we present and evaluate various opportunities that the FDN offers in achieving two main objectives: meeting the SLO requirements and energy efficiency.  All the  opportunities presented in meeting the objectives are evaluated using the implemented \textit{FDNInspector}.

\subsection{Objective 1: Meeting the SLO Requirements}
\label{sec:slo_requriments}

A \textit{Service Level Agreement (SLA)} defines a contract between the provider and the client to meet certain \textit{Service Level Objectives (SLO)}, such as a minimum uptime or a maximum response time. Due to the current homogeneity of nodes in FaaS platforms, it is not possible to scale the function vertically or to provide a specialized machine for its execution.  In the case of heterogeneous target platforms, functions can benefit from the heterogeneity of the underlying platforms and can help in meeting the SLOs. FDN offers multiple opportunities in achieving this and a few of them are presented and evaluated in the following subsections. 

\subsubsection{Function scheduling based on the target platform's performance}
\label{sec:sched_high_hpc}
One method is to always invoke the function on a target platform that has the highest compute capability (and gives the best performance). The \textit{hpc-node-cluster} performed best among all target platforms as shown in Figure~\ref{functions_compare_three_target_platforms}. Therefore, function invocations can always be scheduled on this target platform to meet the SLO, with the assumption that no new target platform is added to the FDN. If a new target platform is added, then first it needs to be benchmarked to analyze it's performance and then accordingly ranked among the target platforms (in this work \textit{hpc-node-cluster}, \textit{old-hpc-node-cluster}, \textit{cloud-cluster}, and \textit{edge-cluster} respectively). Following this, the functions requiring strict SLOs can be scheduled on the target platform with the highest performance. 

% If a new target platform is added, then one needs to conduct an initial benchmarking of it, to understand its compute capacity and 

% Invoking the function always on the highest resources target platform. In the Figure ~\ref{functions_compare_three_target_platforms}, \textit{hpc-node-cluster} performed the best for all the functions. Therefore one can schedule the function invocations always on this target platform for meeting the SLO  with the assumption that no new target platform is added to the FDN. If a new target platform is added, then one needs to conduct an initial benchmarking of it to understand its computational capacity and then rank accordingly all the target platforms (in this work \textit{hpc-node-cluster}, \textit{old-hpc-node-cluster}, \textit{cloud-cluster} and lastly \textit{edge-cluster}) and schedule the function requiring strict SLOs on the highest computational capacity target platform. 
 
\begin{figure}[t]
\centerline{\includegraphics[width=\linewidth]{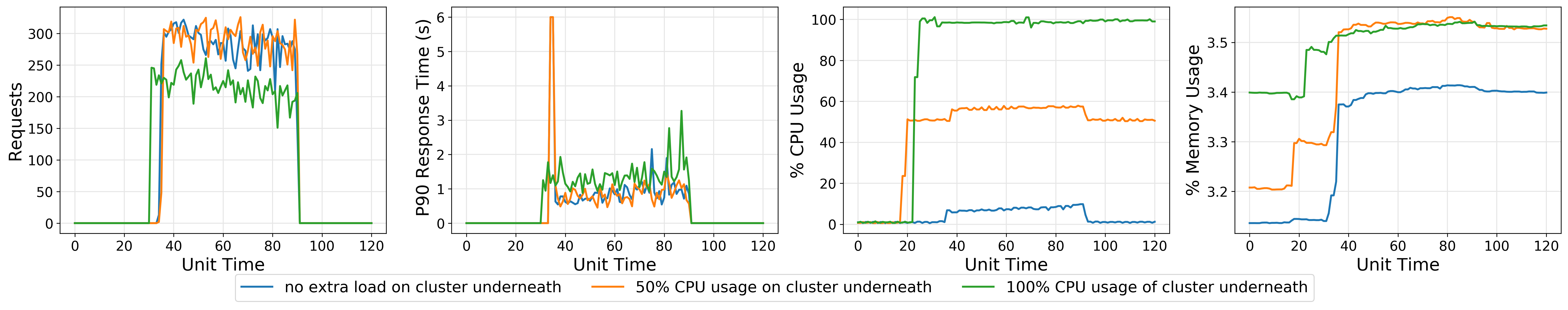}}
\caption{ Performance comparison of \textit{image-processing} function with invocations generated from 40 VUs on the  \textit{old-hpc-node-cluster} in three scenarios: 1) when the cluster is idle, 2) when the target platform has additional 50\% CPU load on it, and 3) when the target platform has additional 100\% CPU load on it. Scheduling the function invocations on a cluster with 100\% CPU load underneath can impact its performance.  
\label{compare_image_process_inwest_load_93_94}}
\end{figure}

\begin{figure}[t]
\centerline{\includegraphics[width=\linewidth]{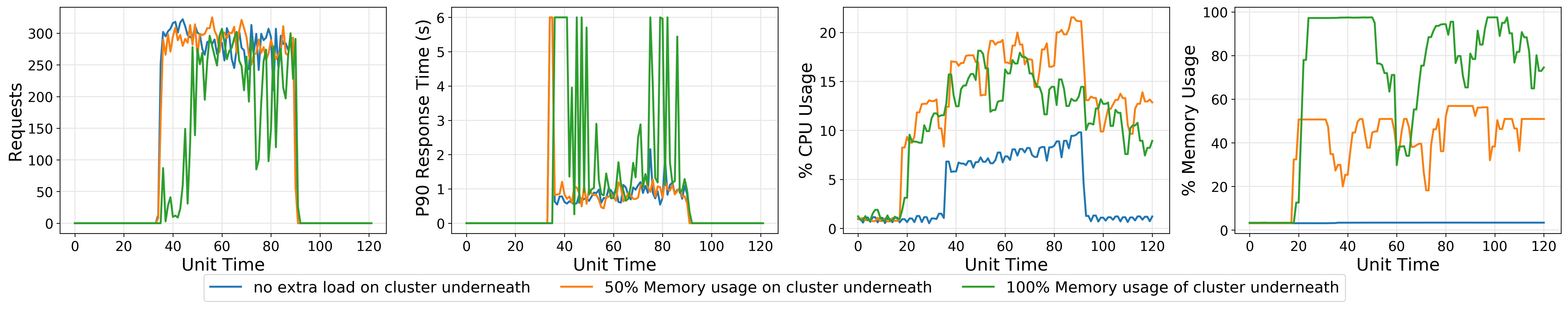}}
\caption{ Performance comparison of \textit{image-processing} function with invocations generated from 40 VUs on the  \textit{old-hpc-node-cluster} in three scenarios: 1) when the cluster is idle, 2) when the target platform has additional 50\% Memory load on it, and 3) when the target platform has additional 100\% Memory load on it. Scheduling the function invocations on a cluster with 100\% Memory load underneath can impact its performance significantly.  
\label{compare_image_process_inwest_load_93_98_mem}}
\end{figure}

\subsubsection{Scheduling based on the target platform's resource utilization}
\label{sec:cluster_usage}
Scheduling the function invocations on the target platform with the highest compute capability will not always lead to the best performance. For example, in the situation in which a workload is already running on the target platform. In this case, scheduling function invocations on it can hamper the performance of both the workloads and result in SLO violations. Therefore, it is important to know the usage of each target platform before scheduling functions on them. Figure~\ref{compare_image_process_inwest_load_93_94} shows the performance comparison of \textit{image-processing} function invocations for 40 VUs on the \textit{old-hpc-node-cluster} across three scenarios:  1) when there is no additional workload on the target platform, 2) when the target platform has an additional 50\% CPU load on it, and 3) when the target platform is fully utilized. Scheduling the function invocations on a target platform with 100\% CPU load leads to a degradation in its performance (the P90 response time approximately increased from 0.8s to 1.5s and the number of requests processed decreased by 100 per unit time). However, for scenario 2, no decrease in performance is seen.

% Not always scheduling the function invocations on the highest computational capacity target platform can result in the best performance. For example, the case of where already some load is running on the target platform and scheduling function invocations on it can hamper the performance of both the workloads. Therefore, one must know the usage of each target platform before scheduling functions on them. Figure~\ref{compare_image_process_inwest_load_93_94} shows the performance comparison of \textit{image-processing} function invocations from 40VUs on the  \textit{old-hpc-node-cluster} in three scenarios:  1) when there is no additional workload on the target platform, 2) when the target platform is having an additional 50\% CPU load on it, and 3) when the target platform is fully computationally utilized. Scheduling the function invocations on a target platform with 100\% CPU load underneath impacted its performance (here P90 response time approximately increased from 0.8s to 1.5s and the number of requests processed decreased by 100 per unit time). However, the target platform having an additional 50\% CPU load does not impact the performance. 

The performance comparison of \textit{image-processing} function for the same three scenarios for 40 VUs on the \textit{old-hpc-node-cluster} with additional load on memory rather than on the CPU is shown in Figure~\ref{compare_image_process_inwest_load_93_98_mem}. When a function is invoked it leads to the creation of function replicas that require a certain amount of memory (in this work 256MB). If the required memory is not available, the performance is decreased as shown in Figure~\ref{compare_image_process_inwest_load_93_98_mem}. Scheduling function invocations on a target platform with 100\% memory load leads to a significant decrease in performance (the P90 response time approximately increased from 0.8s to 6s). However, similar to Figure~\ref{compare_image_process_inwest_load_93_94} invoking functions on a machine with an additional 50\% memory load does not affect performance as there is still free memory available for creating additional function replicas.

Therefore, because of the heterogeneity offered by the FDN, offloading the function invocations from one target platform with high resource utilization to another with a lower value will result in meeting the SLOs. Furthermore, this approach can also be applied for placing or scheduling functions together if they are using complimentary resources. For instance, placing memory-intensive and compute-intensive functions together on the same target platform and placing other compute-intensive functions on different target platforms will lead to optimal utilization of the underlying resources. In this way, the performance of functions will not decrease on simultaneous execution.

\subsubsection{Collaborative execution based on the utilization of different target platforms}
\label{sec:multiple_target_platforms}
\begin{figure}[t]
\centerline{\includegraphics[width=\linewidth]{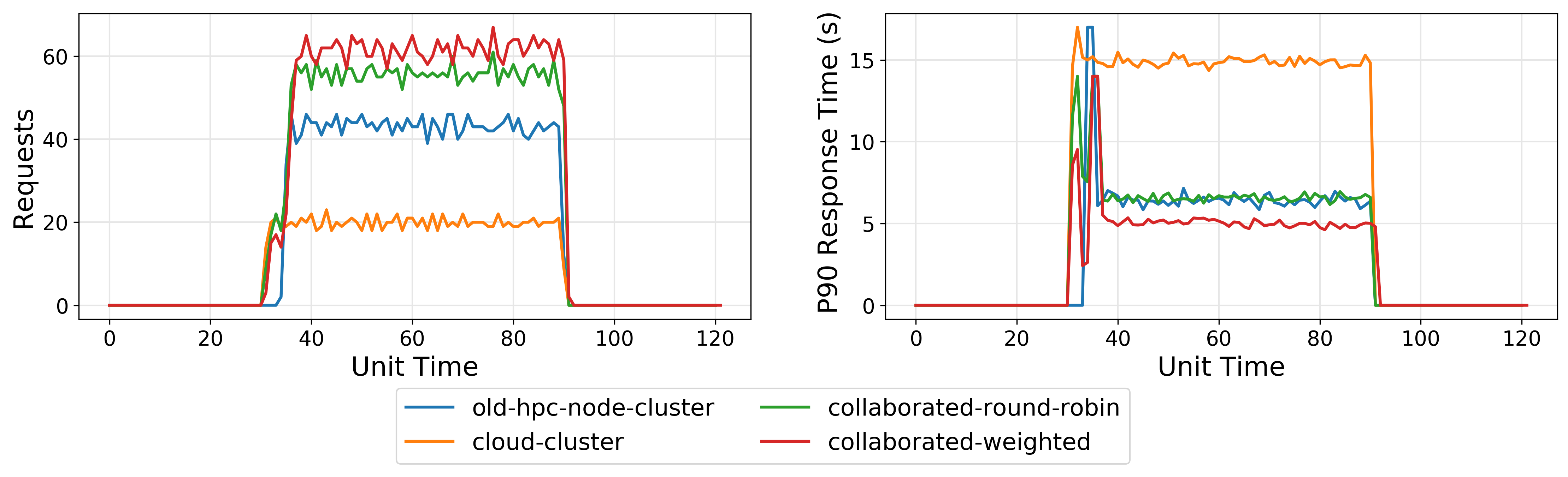}}
\caption{ Performance comparison of \texttt{primes-python} function with invocations generated from 30 VUs on the \textit{old-hpc-node-cluster}, \textit{cloud-cluster}, and when both were collaborated together in round-robin and weighted load balancing manner.  
\label{compare_primes_python_combined_individual_weighted}}
\end{figure}
% Another way of meeting the SLO requirements is by load balancing the function invocations between multiple target platforms. This can even be helpful in the cases where there is some additional load on a target platform and scheduling function invocations on multiple target platforms can prevent from degrading the performance. In this work, we created a NGINX server in between the \textit{old-hpc-node-cluster} and \textit{cloud-cluster} for load balancing the invocations to both clusters in two scenarios (see the Figure~\ref{compare_primes_python_combined_individual_weighted}): 
This method can be helpful in scenarios where there is some additional load on the target platform and scheduling function invocations on multiple target platforms can prevent performance degradation and SLO violations. In this work, we created a NGINX server in between the \textit{old-hpc-node-cluster} and the \textit{cloud-cluster} for collaborating the function invocations to both clusters as shown in Figure~\ref{compare_primes_python_combined_individual_weighted}. We consider two scenarios : 

% Another way of meeting the SLO requirements is by load balancing the function invocations between multiple target platforms. This can even be helpful in the cases where there is some additional load on a target platform and scheduling function invocations on multiple target platforms can prevent from degrading the performance. 

\begin{enumerate}[1.]
 \item \textbf{Round-robin Collaboration}: In this case the function invocations are distributed across both target platforms in a round robin manner.
   \item \textbf{Weighted Collaboration}: The biggest drawback of using the round robin approach is that it assumes target-platforms are similar enough to handle equivalent loads. However, because of heterogeneous target platforms in the FDN, the algorithm has no way to distribute more or less requests to these target platforms based on their resources. As a result, target platforms with less capacity may overload and fail more quickly while capacity on other target platforms remains idle. Therefore, in this case we use weighted collaboration, where function invocations are distributed across the two target platforms based on the weights assigned to each target platform.  In this work, \textit{old-hpc-node-cluster} target platform is assigned a weight of five and \textit{cloud-cluster} of one, which means that out of total six function invocations, five will be invoked on the \textit{old-hpc-node-cluster} and one on the \textit{cloud-cluster}.
%  \item \textbf{Weighted load balance}: The biggest drawback of using the round robin approach is that it assumes that target-platforms are similar enough to handle equivalent loads. However, because of heterogeneous target platforms in FDN, the algorithm has no way to distribute more or less requests to these target platforms based on their resources. As a result, target platforms with less capacity may overload and fail more quickly while capacity on other target platforms lie idle. Therefore in this case the weighted load balancing is used where function invocations are distributed across the two target platforms based on the wieights assigned to each target platform. In this work, \textit{old-hpc-node-cluster} target platform is assigned the weight of 5 and \textit{cloud-cluster} as 1 which means out of total 6 function invocations, 5 will be invoked on \textit{old-hpc-node-cluster} and 1 on \textit{cloud-cluster}. 
 \end{enumerate}

% The \texttt{primes-python} function was deployed on each of the two target platform (\textit{old-hpc-node-cluster} and \textit{cloud-cluster}) and a load of function invocations from 30 VUs was generated on the  \textit{old-hpc-node-cluster}, \textit{cloud-cluster},  when both are combined together with round-robin load balancing and, lastly when both are combined with weighted load balancing. Figure~\ref{compare_primes_python_combined_individual_weighted} shows the performance comparison for all these 4 cases. \textit{combined-round-robin-cluster} was able to achieve significantly better performance from the alone \textit{cloud-cluster} (increased from 20 per time to 55 per unit time and at a lower response time 6 seconds per request). Also, comparing it to alone \textit{old-hpc-node-cluster} target platform's performance, it was able to serve higher number of requests with approximately similar response time. Furthermore,  \textit{combined-weighted-cluster} outclassed all the others by serving more number of requests (approximately 60 per unit time) and at a lower response time (5 seconds per request). 
We deploy the \texttt{primes-python} function on the two target platforms \textit{old-hpc-node-cluster} and \textit{cloud-cluster}. To demonstrate the benefits of collaborative function invocations between multiple target platforms, we consider four scenarios. In scenarios 1 and 2, all functions are invoked exclusively on the  \textit{old-hpc-node-cluster} and \textit{cloud-cluster} respectively. For scenarios 3 and 4, the two target platforms are collaborated together with round robin and weighted manner. For all scenarios, we generate a load of 30 VUs. The performance comparison for all the four scenarios is shown in Figure~\ref{compare_primes_python_combined_individual_weighted}. We observe a significant increase in performance when the two platforms are collaborated with round robin manner as compared to when the functions are invoked exclusively on the \textit{cloud-cluster}. In this case, the number of requests processed increased from 20 to 55 per unit time with a lower P90 response time of six seconds per request. Moreover when compared to scenario 1, the number of requests served were higher in scenario 3 with approximately the same P90 response time. We observe the best performance in scenario 4, i.e., with weighted collaboration. In this case, 60 requests per unit time were served with a response time of five seconds per request.

Collaboration between multiple heterogeneous target platforms in the FDN is a method to overcome the shortcomings of individual target platforms. Additionally, this mechanism can also be used to reduce the cold-start problem. This can be done by keeping a low resource cluster always warm and directing initial function invocations to it and later using weighted collaboration between other target platforms. Moreover, it is also possible to create a dynamic rule inside the load balancer that checks for the warm target platform and directs the initial function invocations to it leading to a overall better performance.
% Load balancing between multiple heterogeneous target platforms in FDN allows to overcome the shortcomings of individual target platforms. Additionally, this mechanism can also be used to reduce the cold-start problem, where one can keep a low resources cluster always warm and load balance the initial function invocation requests to it and later on apply the weighted distributes to other target platforms. Also, one can create a dynamic rule inside the load balance to check for the warm target platform and directs the requests to it which will allow to achieve a better performance. 

\subsubsection{Data Localisation}
\label{sec:data_locale}
\begin{figure}[t]
\centerline{\includegraphics[width=\linewidth]{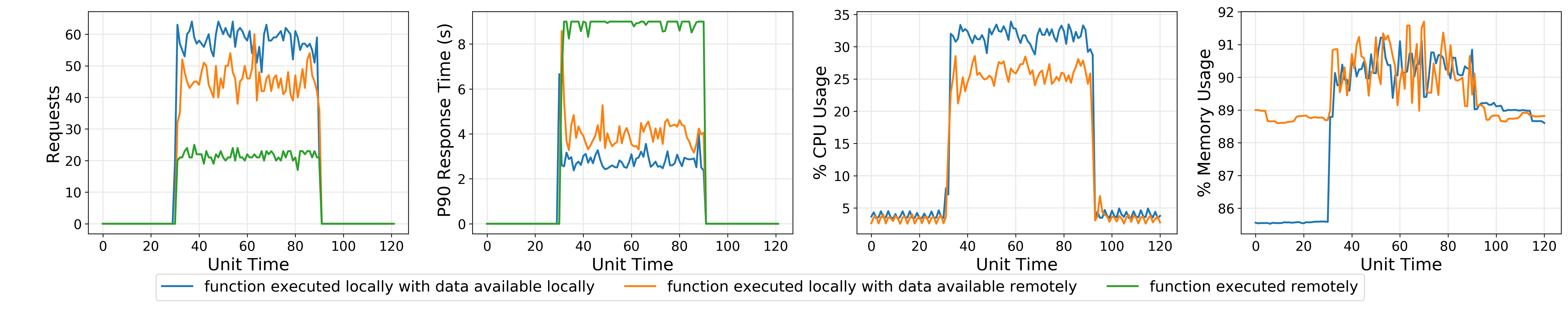}}
\caption{ Performance comparison of \textit{image-processing} function with invocations generated from 20 VUs on the  \textit{cloud-cluster} when the data is available locally and remote.  
\label{compare_image_process_data_localisatipn}}
\end{figure}

Although functions in FaaS are stateless, changes in state and look-ups require frequent access to databases and object storages. Current platforms do not take into account the data access behaviour of functions while scheduling. This leads to longer execution times and a violation of the SLO requirements. To demonstrate this, we hosted 2 MinIO~\cite{MinIO:online} instances: one locally on the target platform and another remotely on the Google Compute Platform (GCP) in \texttt{us-east} region. MinIO is an object store, which can store unstructured data such as photos, videos, log files, backups and container images. Following this, we evaluated the performance of the \textit{image-processing} function, that takes an image from the two MinIO instances and performs different operations on them. For our experiments, we used the \textit{cloud-cluster} with function invocations for 20 VUs for accessing the data from the two MinIO instances and \textit{google-cloud-cluster} with same number of function invocations for showcasing the performance when the function is scheduled closer to the remote data storage. Figure~\ref{compare_image_process_data_localisatipn} shows the performance comparison for these three scenarios. The \textit{cloud-cluster} with function invocations accessing the local MinIO instance was able to serve more requests (approx. 60 per unit time) than when accessing the remote MinIO instance (approx. 45 per unit time) and at a lower P90 response time (three seconds per request than four seconds per request in case of the remote MinIO instance). Executing the function on the \textit{google-cloud-cluster} performed worst with 20 requests per unit time at P90 response time of 8.5 seconds. This can be attributed towards the inability of the GCFs to handle compute intensive functions and also to the large latency caused by difference in the regions from where the request is executed (in Germany) and where the request is handled (in us-east region).  
% Currently, functions are stateless and the state changes and look-ups require frequent access to databases and object storages, current platforms do not take into account the data access behavior of functions for scheduling the functions. This can results in longer execution latency and violation of the SLO requirements. 
% We hosted 2 MinIO instances: one locally on the target platform and another remotely on the AWS cloud in the us-east region. MinIO is an object store, which can store unstructured data such as photos, videos, log files, backups and container images. Following this, we evaluated the performance of the \textit{image-processing} function, which takes an image from MinIO and performs different operations for these two different MinIO instances using the \textit{cloud-cluster} as the target platform with the function invocations from 20 VUs. 

% Figure~\ref{compare_image_process_data_localisatipn} shows the performance comparison results for these two scenarios. The \textit{cloud-cluster} with function invocations to the MinIO instance running locally was able to serve more requests (approx. 60 per unit time) than the remote MinIO instance (approx. 45 per unit time) and at a lower response time (3 seconds per request than 4 seconds per request in case of remote MinIO instance). 

Migrating data closer to the target platform can significantly reduce the access latency. Hence, adaptive data management is a key part of the FDN in meeting the SLOs. For instance, data required for training a neural network can be migrated to a high-performance target platform. This will reduce the data access latency leading to a decrease in training time.  Furthermore, a subset of the data can be migrated to edge-cluster for low-latency machine learning model inference. Additionally, placement of the functions closer to data location can provide an another way of achieving a lower access latency. However, in our experiments when we executed the function on \textit{google-cloud-cluster} which is closer to data performed worst due to the large difference in between the execution and processing locations. Nevertheless, one can use such a strategy to handle large function requests when the local cluster doesn't have enough resources for handling the requests.  

% For example data required for training a neural network can be migrated to the high resources target platform which will reduce data access latency and hence training can be done at a faster rate than before. Furthermore, a subset of the data can be migrated to edge-cluster for low-latency machine learning model inference. 

\subsection{Objective 2: Energy Efficiency}
\label{sec:energy_efficieny}
Another important objective which is highlighted in the FDN is providing energy efficiency for certain amount of workloads due to the availability of resource-constrained target platforms like our \textit{edge-cluster}. This cluster is made up from Nvidia's Jetson Nano edge devices and consumes significantly less energy than the other target platforms. 

To obtain power measurements for the Jetson Nano edge devices, we utilize the inbuilt power monitors~\cite{NVIDIADe72:online} that measure power consumption for different supply rails. Specifically, we measure the power consumption for the rail \texttt{POM\_5V\_CPU}. On the other hand, the power consumed by the \textit{hpc-node-cluster} is obtained through running average power limit (RAPL) counters \texttt{PKG0} and \texttt{PKG1} for the two sockets respectively. It is important to note that for all experiments we measure the CPU power consumption and average the power values over five runs of the same experiment. We evaluate the energy consumed by \textit{edge-cluster} and \textit{hpc-node-cluster} when a load of 400 requests per second from 40 VUs is invoked on the function \texttt{JSON-loads} deployed on each one of them. We calculate the energy consumed by multiplying the average power with the duration of the experiment. Although, the P90 response time (6.32s) is  higher for the \textit{edge-cluster} as compared to \textit{hpc-node-cluster} (2.3s), the total number of requests served is same for both target platforms (400 requests per second). Therefore, if a client has a SLO P90 response time of seven seconds then both target platforms can be used for meeting it for this workload. However, there is a significant difference in the CPU energy consumption of the target platforms as shown in Table~\ref{energy_consumed}. For the \textit{edge-cluster}, we obtain a total CPU energy consumption of around \textbf{2647.2 Joules} as compared to \textbf{44645.64 Joules} for the \textit{hpc-node-cluster}. Table~\ref{energy_consumed} also shows the individual CPU power consumption with and without workload for each node in \textit{edge-cluster} and for each socket in \textit{hpc-node-cluster}. Clearly, choosing  \textit{edge-cluster} as the target platform for this small workload saves a lot of energy. Automatically placing the functions on the low energy consumption target platform based on the workload is part of the FDN. 
% We evaluated the energy consumption by \textit{edge-cluster} and \textit{hpc-node-cluster} when a load of 40 requests per second using 5 VUs is invoked on the function \texttt{nodeinfo} deployed on each one of them. Although, the requests response time (100ms per request) on the \textit{edge-cluster} is slightly higher than that of \textit{hpc-node-cluster} (60ms per request) but the overall number of requests  served is same (40 requests per second) for both. Therefore, if a client has a SLO response time of 1s per request then both the target platforms can be used for meeting it at this very load. However, the power consumption of each one them as shown in the Table~\ref{energy_consumed} differs significantly with the  \textit{edge-cluster}'s around \textbf{1612.2 Joules}  and while that of \textit{hpc-node-cluster} is \textbf{66150.1 Joules}. Table~\ref{energy_consumed} also lists the power consumption without and with workload for each node in \textit{edge-cluster} and each socket in \textit{hpc-node-cluster} target platforms.   Clearly, choosing  \textit{edge-cluster} target platform for this small load saves a lot of energy. Automatically placing the functions on the low energy consumption target platform based on the load is part of the FDN.  

%Although the response time differs from two clusters 

%The Intel Westmere EX processor does not support reading power data

\begin{center}
\begin{table}[t]%
\centering
\caption{Total energy consumption for \textit{edge-cluster} and \textit{hpc-node-cluster} target platforms when a load of 400 requests per second using 40 VUs is invoked on the function \texttt{JSON-loads}.  \label{energy_consumed}}%
\begin{threeparttable}
\begin{tabular*}{500pt}{@{\extracolsep\fill}lcccD{.}{.}{3}c@{\extracolsep\fill}}
\toprule
&\multicolumn{3}{@{}c@{}@{}}{\textbf{edge-cluster}} & \multicolumn{2}{@{}c@{}}{\textbf{hpc-node-cluster}} \\\cmidrule{2-4}\cmidrule{5-6}
\textbf{} & \textbf{Node 1}  & \textbf{Node 2}  & \textbf{Node 3} & \multicolumn{1}{@{}l@{}}{\textbf{Socket 0}}  & \textbf{Socket 1}   \\
\midrule
\textbf{CPU power consumption without workload (W)} &  0.457 & 0.390 & 0.489 & 30.6757 & 29.566   \\
\textbf{CPU power consumption with workload (W)} &  1.414 & 2.046 & 0.952 & 37.399 & 37.01   \\

%\textbf{CPU power consumption without workload (W)} &  1.414 & 1.403 & 1.481 & 54.930 & 53.800   \\
%\textbf{CPU power consumption with workload (W)} &  2.722 & 3.574 & 2.270 & 55.860 & 54.390   \\
\midrule
\textbf{Total CPU energy consumption (J)} &  \multicolumn{3}{@{}c@{}@{}}{\textbf{2647.2}}  & \multicolumn{2}{@{}c@{}}{\textbf{44645.64}}  \\
\bottomrule
\end{tabular*}
\end{threeparttable}
\end{table}
\end{center}

\section{Discussion}
\label{sec:discussion}
% In this section, we discuss a few other aspects that can be taken into account when scheduling functions in FDN. These include the role of underneath FaaS platforms, optimization of application based on the underneath h/w specifications, and the impact of runtime environment.

In this section, we discuss a few other opportunities offered by the FDN that can be used for achieving various objectives.  %These include the role of the underneath FaaS platforms, optimizing applications based on the underlying hardware specifications, and the impact of the language runtime environment.

\subsection{Heterogeneity among FaaS platforms}
\label{sec:hetrogenous_faas}
We have used three different FaaS platforms in this work (OpenWhisk, OpenFaaS and GCFs) but there are several others which are available including the ones offered by the public cloud providers like AWS lambda function from Amazon, and Azure functions from Microsoft. One can integrate all these together into the FDN. However, mapping the metrics from all these platforms to a common metric so that the FDN can make decisions is challenging due to differences in their semantics, aggregation, and measurement. 

Mature platforms like OpenWhisk use optimized caching and distinguish between cold, prewarm and warm containers to address the cold-start problem~\cite{mohan2019agile}. Prewarm containers are containers that already have the runtime environment for an action set up. For example, when OpenWhisk's algorithm anticipates Node.js based actions, it will start preparing generic Node.js containers, which reduces most of the cold-start time. When an action is executed very frequently, OpenWhisk will detect that and keep its containers warm. Warm containers are containers where the action is already initialized and ready to be run at any time. On the other hand, OpenFaaS does not have the concept of warm and pre-warm containers as a result this can affect the performance on the target platform when using it. OpenFaaS like OpenWhisk does support the option to scale to zero and hence save money on idle resources. Additionally, OpenFaaS provides support for low-end edge devices with ARM processors and therefore is a clear candidate for usage on edge target platforms. Public cloud providers FaaS platforms provide an advantage of executing the functions globally in any region of the world and also have large scaling capabilities. 

% FDN offers these heterogeneity between FaaS platforms by which multiple system architecture devices like android devices based target platform can as well be included as part of this and then the functions can be deployed on them. Further, using the FaaS platform optimized for a certain system architecture devices and using that combination as target platform can help in achieving a higher performance and better SLOs like tinyFaaS platform optimized for edge environments, which the application developers can also exploit. 
FDN offers heterogeneity between FaaS platforms through which multiple devices with different system architectures such as android phones can be integrated into it. Furthermore, using the FaaS platform optimized for certain system architectures such as \textit{tinyFaaS} for edge devices can lead to a higher performance and better SLOs. This can be exploited by the application developers.

% Further, mature platforms like OpenWhisk uses optimized caching and distinguishes between cold, prewarm and warm containers to address the cold-start problem[51]. Prewarm containers are containers that already have the run environment for an action set up. For example, when OpenWhisk's algorithm anticipates Node.js based actions, it will start preparing generic Node.js containers, which reduces most of the cold-start time. When an action is executed very frequently, OpenWhisk will detect that and keep its containers warm. Warm containers are containers where the action is already initialized and ready to be run at any time. On the other hand, OpenFaaS does not have the concept of warm and pre-warm containers as a result this can affect the performance on the target platform using OpenFaaS. OpenFaaS like OpenWhisk does support the option to scale to zero and hence save money on idle resources. Additionaly, OpenFaaS does provide the support for low-end edge devices with ARM processor and therefore makes a clear candidate for running it on edge target platforms. Therefore, this heterogeneity of the FaaS platforms should be taken into account when scheduling the functions in FDN.
\subsection{Application optimization based on the underlying hardware}
\label{sec:sys_arch}
Due to the availability of heterogeneous target platforms in FDN, application function developers can optimize their code to use specialized hardware like GPUs or specialized processor features like SIMD/AVX for running their functions. For designing such applications, hints or recommendations on which target platform the function will be scheduled by the FDN can be provided to the developer which will allow developers to target code for specific hardware features, to provide innovative hardware/software co-design. Moreover, application developers could provide functions using a high-level domain specific language, and the FDN can automatically compile these functions to the most cost-effective target platform based on the user specified SLOs. 

Furthermore, conventional enabling technologies for ML at edge networks require personal data to be shared with external parties, e.g., edge servers. Recently, in light of growing data privacy concerns, the concept of Federated Learning (FL) has been introduced~\cite{kairouz2019advances}. In FL, end devices use their local data to train an ML model required by the server. The end devices then send the model updates rather than raw data to the server for aggregation. FL can serve as an enabling technology in edge networks. However, in a large-scale and complex mobile edge network, heterogeneous devices with varying constraints are involved which raises challenges of resource allocation in the implementation of FL at scale. FDN having heterogeneous target platforms including the edge can be used to provide automatic resource allocation and function scheduling for FL based applications. 

\subsection{Function Composition}

Functions can also be chained together into sequences where chained functions use the output of the preceding function as input. In AWS lambda platform these are called step functions~\cite{AWSStep:online}. AWS lambda charges an additional cost for each transition from one function to another~\cite{AWSStepPricing:online}. Therefore, in such cases, in order to reduce the number of state transitions to make the overall deployment cost-efficient without violating the SLOs, multiple functions can be composed together. Moreover, functions with different functionalities can be composed together into a larger function to meet user requests when serving requests from a single function is not possible. For example, if the output parameters of one function can be used as the input parameters of another function, these two functions can be connected as a new function with input parameters that are the same as the input parameters of the first single function and output parameters that are the same as the output parameters of the second function. This new function is called a composed function, and the elemental functions are referred to as the member functions. Deploying these member functions together on a target platform having higher compute capability can result in higher QoS of the overall application. One way of achieving this in FDN is by deploying member functions together within a kubernetes pod, and then the two functions will always be deployed together, resulting in lower number of transitions and cost reduction. In~\cite{kratzke2018brief}, the author mentioned about the problem of double-spending with function composition, where a serverless function (composer function) whose purpose is to just call other serverless functions is also billed to the user although only the called functions are consuming the resources. FDN can recognize such composer functions and automatically schedule them to the low-resources target platforms to reduce the overall cost.

%\subsection{Role of language runtime}
%\label{sec:sys_arch}

%\subsection{Explotation for other objectives}
%\label{sec:sys_arch}
%system level, 

% \subsection{composability}
% chaining of functions
% Functions can also be chained together into sequences where chained functions use the output of the preceding function as input.

% Each of the FaaS platform does have an official list of the language runtimes they support, however, every FaaS platform even supports custom Docker runtimes, which means the officially supported languages are merely more convenient to use. Although we have not measured the performance impact of different language runtimes but we believe different language runtimes can affect the quality of service of the functions based on some previous works~\cite{wg2018cncf}. Therefore when scheduling functions on target platforms in the FDN, language runtime should also be taken into account. 
\section{Related Work}
\label{sec:related_work}

Researchers have already identified the limitations of current serverless platforms, such as no control over specifying additional hardware resources like the required number of CPUs, GPUs, or other types of accelerators for the functions, and inefficient communication patterns between functions because of the data access latency~\cite{hendrickson2016serverless,baldini2017serverless, hellerstein2018serverless}. Jonas et al.~\cite{jonas2019cloud} suggest some improvements and workarounds which can be adopted to overcome these limitations. Since the FDN targets heterogeneous platforms, it overcomes these limitations by taking into account the computational (CPUs, GPUs etc.) and data requirements (remote or local data availability) of the function and then schedules the function automatically on the right target platform. In this process, migration of data closer to the function can also take place if there is a higher data access latency. Furthermore, Shahrad et al.~\cite{shahrad2019architectural} studied the architectural implications of serverless computing and pointed out that exploitation of system architectural features like temporal locality and reuse are hampered by the short function runtimes in FaaS. In the FDN, application function deployment hints regarding deployment target platforms of the functions will be provided to the user from which the developer can exploit the system architectural features to optimize the application and achieve higher SLOs.

In the following paragraphs, we present prior work from three aspects: (i) heterogeneity in public cloud providers FaaS platforms performance and how FDN can take advantage of this, (ii)  FaaS for HPC and how FDN can be advantageous for HPC workloads, (iii) FaaS for edge devices, and (iv) different strategies for coordination among heterogeneous platforms and how FDN strategies differs from them.  

% Researchers have already pointed out towards the limitations of the current serverless platforms, such as no control over specifying the additional hardware resources like the number of CPUs, GPUs, or other types of accelerators requirement for the functions and inefficient communication patterns between functions because of the data access latency~\cite{hendrickson2016serverless,baldini2017serverless, hellerstein2018serverless, jonas2019cloud}. Jonas et al. have suggested some improvements and workarounds which can be adopted to overcome these limitations~\cite{jonas2019cloud}. Furthermore, Shahrad et al. studied the architectural implications of serverless computing and pointed out that exploitation of system architectural features like  temporal locality and reuse are hampered by the short function runtimes in FaaS~\cite{shahrad2019architectural}. Since FDN is targeted for heterogeneous platforms, therefore it overcomes these limitations by taking into the consideration the computational (CPUs, GPUs etc.) and data requirements (remote or local data availability) of the function and then schedule the function automatically on the right target platform. In this process, migration of data closer to the function can also take place if there is a higher data access latency. In addition, FDN is not limited to short running functions, therefore long running heterogeneous resources requirement functions which can exploit the necessary architectural features can easily be scheduled on it.  

FaaSProfiler~\cite{shahrad2019architectural} is the first to take a bottom-up approach in analyzing the architectural implication to unwrap the server-level overheads in the FaaS model. They analyzed the difference between native and in-FaaS function execution and calculated the additional server-level overheads like computational overheads, memory consumption, bandwidth usage, and management overheads like orchestration, queuing, scheduling, and power consumed. Furthermore, Lee et al.~\cite{lee2018evaluation} compared the performance of various serverless computing environments offered by public cloud providers by showcasing the results of throughput, network bandwidth, file I/O and compute performance regarding the concurrent function invocations. L.Wang et al.~\cite{wang2018peeking} performed an in-depth study of resource management and performance isolation with three popular serverless computing providers: AWS Lambda, Azure Functions, and Google Cloud Functions. Their analysis demonstrates a reasonable difference in performance between the FaaS platforms and states that azure functions use different types of VMs hosts and 55\% of the time a function instance runs on a VM with debased performance. They have also stated that on Azure the functions host VMs can have 1, 2 or 4 vCPUs. Additionally, K. Figiela et al.~\cite{figiela2018performance} developed a cloud function benchmarking framework. CPU intensive functions were deployed in major cloud providers FaaS platforms. The authors observe fluctuation in response time duration based on the different underlying hardware, runtime systems, and resource management. These observations showcase the heterogeneity in the performance and resources availabilities from the public cloud FaaS offerings. Thus, FDN across these public FaaS platforms can provide a way for enabling the scheduling of the functions on them by delivering the function in a right platform based on its requirements in such a way that the performance is adhered to the defined SLOs at the lowest cost. 

Lynn et al.~\cite{8241104} studies seven different public serverless platforms including, AWS Lambda, Google Cloud Functions, and Microsoft Azure Functions, to showcase that serverless computing can be applied to a wide range of use cases.
Serverless computing is highly relevant for scientific applications, especially in conjunction with HPC capabilities~\cite{jonas2017occupy, 2017arXiv170808028F}.
PyWren~\cite{jonas2017occupy} utilized an external ad-hoc orchestrator to share state and synchronize parallel execution of functions in simple map-reduce applications. There has also been some work to enhance the function startup latencies such as SAND~\cite{215935} in which the authors utilized application-level sandboxing, and a hierarchical message bus for achieving shorter startup delays and efficient resource usage. McGrath et al. in~\cite{7979855} proposed a queuing scheme with workers in which function containers that can be reused are put into warm queues and workers where new containers need to be created are put into cold queues. Splillner et. al.~\cite{10.1007/978-3-319-73353-1_11} demonstrated that FaaS cloud model can be used for different
HPC batch workloads, such as, calculating the value of $\pi$, image face detection, password cracking, and weather forecasting. Malla et al.~\cite{Malla2020HPCIT} compared Google Cloud Functions with Google Compute Engine  in terms of cost and performance for a HPC workload. They found that FaaS can be 14\% to 40\% less expensive than IaaS for the same level of performance, but, performance of FaaS exhibits higher variation due on-demand CPUs allocation by the cloud service providers. Based on these observations we have integrated FDN to HPC nodes cluster platform. Furthermore, FDN provides the option of scheduling the HPC based workload function to more performant HPC nodes cluster platform or to highly available and scalable public cloud FaaS platforms. The decision to choose a platform can be made based on the user requirements such as performance vs cost. In our previous work, we used similar approach for achieving federated learning using heterogeneous FaaS platforms ~\cite{10.1145/3429880.3430100}.

The first documented efforts for bringing serverless capabilities to the edge came from industry with the introduction of  AWS Lambda@Edge\footnote{http://docs.aws.amazon.com/lambda/latest/dg/lambda-edge.html} that allows one to explicitly deploy lambda functions to edge locations. This is then used within the IoT Greengrass system of Amazon~\cite{AWSIoTGr10:online}. It allows to integrate edge devices with cloud resources in an IoT platform and application Lambda functions running on it are deployed to the edge computers. Baresi et al.~\cite{10.1007/978-3-319-67262-5_15} propose a serverless model for Multi-Access Edge Computing (MEC). They provide a broader range of application scenarios along with optimizations that compose a serverless edge platform.
KubeEdge~\cite{xiong2018extend} is an open source system extending native containerized application orchestration and device management to hosts at the edge. These frameworks focus on executing the applications only on the edge by extending cloud based FaaS platforms on the edge. Pfandzelter et al.~\cite{pfandzelter2020tinyfaas} highlight the problem of running cloud based FaaS platforms on the edge and introduce a new FaaS platform called tinyFaaS for edge environments. FDN includes \textit{edge-cluster} platform, allowing the opportunity of scheduling the functions closer to the user and hence providing a better performance. Furthermore, FDN allows multiple instances of same function to coexist across multiple heterogeneous platforms, thus providing a way for handling function invocations from various opportunistic requirements.

With respect to these works, our proposed FDN provides a way for cooperation among various heterogeneous platforms for increasing the performance, robustness and scalability of these platforms. To share resources efficiently for multiple tasks in the cloud, a game-theoretic approach is introduced by Freeman et al.~\cite{freeman2018dynamic}. Designed for latency critical applications, PARTIES~\cite{chen2019parties} presents an online learning approach to efficiently allocate fine-grained resources such as memory bandwidth and last level caches without QoS degradation. Delimitrou et al.~\cite{delimitrou2014quasar} present a cooperative filter based approach to assign a workload to the most appropriate hardware configurations. Satyanarayanan et al.~\cite{satyanarayanan2009case} propose an edge computing approach to offload computation from mobile devices to the network edge using virtual machine (VM) based cloudlets. In fog and edge~\cite{skarlat2016resource, ostberg2017reliable} computing, a considerable amount of research work has also been done for developing methods for resource provisioning and management. Also, there have been studies on integrating edge and cloud computing for allowing the deployment of services on the resource-constrained edge devices and offloading compute-intensive parts to the cloud~\cite{villari2016osmotic, tong2016hierarchical, ifogsim, chen_efficient}. Although the different proposed approaches for resource provisioning show promising results in traditional computing environments, they have not been evaluated and extended for the heterogeneous collection of target platforms in FDN especially involving HPC systems. Bermbach et al.~\cite{auction} have a very particular auction-based approach in which application developers bid on resources fog nodes to make a local decision about which functions to offload while maximizing revenue. It requires no centralized coordination and focuses on maximizing the earnings for the infrastructure provider. On the other hand, there is no guarantee for the user that its function will be executed. Our approach within FDN is designed to have a central coordination point and focuses on the fast response to the user. Hellerstein et al.~\cite{hellerstein2018serverless} describe FaaS as a data-shipping architecture in the sense that it still ships data to code rather than shipping code to data and see it as perhaps the biggest shortcoming of FaaS platforms. The approach of fluid code and data placement, described as \textit{stepping forward to the future}, is the suggested solution to the problem previously mentioned by which the platform would physically colocate certain code and data. Based on this approach, we designed the data migration and function placement strategies in FDN.

To the best of the authors knowledge there has not been any work which involves using heterogeneous platforms Cloud, edge and HPC together for achieving different objectives in a serverless manner.

%We primarily have used Python and Node.js based runtimes for creating the functions in OpenWhisk, where as in OpenFaaS we created our own docker image which can support the underlying ARM processor for different functions. These runtimes and languages are chosen due to their wide support across every major platform. Each FaaS platform provides an official list of the language runtimes it supports. However, it is also possible to use custom Docker runtimes in every FaaS platform. This means that using officially supported runtimes is merely more convenient. Although we have not measured the performance impact of different language runtimes but we believe different language runtimes can affect the quality of service of the functions based on some previous works~\cite{wg2018cncf}. Therefore when scheduling functions on target platforms in the FDN, language runtime should also be taken into account.  

%Integrating serverless on the computing continuum brings another set of challenges~\cite{nastic2018towards}. Due to resources constraints in edge devices not all the serverless platforms can run on them. In~\cite{8817155} four open source serverless frameworks, namely, Kubeless, Apache OpenWhisk, OpenFaaS, Knative are evaluated on resource-constrained edge devices. Furthermore, Pfandzelter et al.~\cite{pfandzelter2020tinyfaas} highlight the problem of running cloud based FaaS platforms on the edge and introduce a new FaaS platform called tinyFaaS for edge environments. 

\section{Threats to validity}
\label{sec:threats_to_validity}

In this section, we discuss potential threats to validity for replicability and reliability of the study, and external validity.

There are two threats towards the replicability and reliability of the study. 
The first one lies in the type of the systems used in this study as experimental target platforms. These systems with the presented configurations may not be publicly available with everyone and hence presents a threat towards the replication of the presented results. However, even if the systems similar to the systems with the presented configurations are used, the authors believe that the drawn conclusions would still be true. Furthermore, the presented study showcase that different heterogeneous platforms provide different opportunities for scheduling functions across the platforms.  Secondly, in this study we can see a potential risk of confirmation bias towards the reliability of the study where we try to to confirm our assumptions. This risk was mitigated by checking ourselves to make sure that we do not have any preference in regard to the outcome. The whole research process is conducted using open source tools along with the standard benchmarks and is made transparent, from how we gathered data, designed the tool and conducted our performance evaluations. Additionally, we open source our designed tool and all the collected data. 

There are two major threats to the external validity of the study. The first one lies in the limitation of the benchmarks used in the work. The benchmarks used are smaller than the complex industrial FaaS applications and do not involve various public cloud service providers BaaS services. The second one lies in the amount of the user workload generated for the benchmarking. The generated user workload may be simpler and smaller than the real workload and represents only a limited part of different types of possible workloads. Thus it is not clear whether the work can be effectively applied for much larger industrial applications and to more complex and real user workloads. Furthermore, the drawn conclusions and opportunities presented in this study may change with the change in the type of platforms used for the evaluations and thus can not be generalized for all the platforms.

%Selection bias.The representativeness of our selected studies isarguably one of the main threats to this study. We used a multi-stage process (see Section 4.1) with sources originating fromdifferentsearchstrategies.Initialmanualsearchbasedonexistingacademic literature collections allowed us to fine-tune the querystring for database searches against 7 well-established electronicresearch databases. We optimize our search string for more in-formal grey literature and query 5 search engines specializing ingeneral-purpose search, social search, and developer-communitysearch. Additionally, our complementary search strategies aimto discover studies that were recently published, found in theother context (i.e., academic vs. grey), or spotted through moreexploratory search (e.g., a looser adaptation of search terms) 

\section{Conclusion}
\label{sec:conclusion}
Due to the current limitations of serverless computing for applications which are highly dynamic in their structure and computational requirements, we introduced the Function Delivery Network (FDN), a network of distributed  heterogeneous target platforms enabling the automatic scheduling of heterogeneous functions to target platforms based on their computational and data requirements. Additionally, the concept of Function Delivery Network (FDN) was evaluated using five distributed target platforms having different computational capabilities (ranging from small edge servers to high-end HPC based machines) for achieving two goals: SLO requirements and energy efficiency using \textit{FDNInspector}, a tool for benchmarking distributed FaaS based target platforms. 
It was found that scheduling function invocations to the high-performance target platform leads to a higher QoS in most cases. However, in the scenario where the target platform's resources are already being used, scheduling functions on it can lead to a degradation in QoS of the application. Therefore, it is important to consider the resource-usage of the target platform before scheduling functions on it. Moreover, collaborating the function invocations between the multiple target platforms can lead to a higher QoS as compared to scenarios where functions are exclusively invoked on individual target platforms. Migrating data closer to the target platform can also significantly reduce the data access latency. We showcase that such opportunities offered by the FDN can help in meeting the SLO requirements. Finally, using an edge-based target platform can achieve significantly lower energy consumption. In this work, we showed that by using an edge-based target platform the overall energy consumption is reduced by 17x as compared to scheduling it on a high-end target platform, without violating the SLO requirements .

% One of our future work is to complete the implementation of the Function Delivery Network and showcase its use for the various dynamic heterogeneous applications such as Federated Learning. In addition, we plan to compare the performance of FDN with a public cloud provider FaaS platform such as AWS Lambda. Also, including AWS lambda as one of the target platform in the FDN is another perspective future scope.    

In the future, we plan to complete the implementation of the Function Delivery Network and demonstrate its use for the various dynamic heterogeneous applications such as Federated Learning. In addition, integrating AWS lambda as one of the target platforms in FDN is another perspective future scope. 

\section*{Acknowledgments}
This work was supported by the funding of the German Federal Ministry of Education and Research (BMBF) in the scope of the Software Campus program. Google Cloud credits were provided by the Google Cloud Platform research credits. We thank the anonymous reviewers for their constructive reviews to improve this work and inspire future work.
%\nocite{*}% Show all bib entries - both cited and uncited; comment this line to view only cited bib entries;
% \bibliographystyle{WileyNJD-v2}
\bibliography{wileyNJD-AMA}%

\end{document}